\documentclass{article}
\usepackage[utf8]{inputenc}
\usepackage{amsmath}
\usepackage{color}
\usepackage{hyperref}
\usepackage{amsmath, amsfonts, amsthm, amssymb}
\usepackage{mathtools}
\usepackage{dsfont}

\usepackage{commath}
\usepackage{verbatim}
\usepackage[export]{adjustbox}
\usepackage[toc,page]{appendix}
\usepackage{stmaryrd} 
\usepackage{fancyhdr}
\usepackage{bbm}
\usepackage{graphicx}
\usepackage{algpseudocode}
\usepackage{graphics}
\usepackage{tikz}
\usepackage{caption}
\usepackage{subcaption}
\usepackage{booktabs}
\usepackage{array}
\usepackage{comment}
\usepackage{todonotes}
\usepackage{tcolorbox}
\usepackage{float}
\usepackage{algorithm}
\usepackage{algpseudocode}
\usepackage{tikz}
\usepackage{multirow,array}
\usepackage{color}
\usepackage{listings}
\usepackage[backend=biber, style=nature, natbib=true, bibstyle=numeric, isbn=false, doi=true, sorting=nyt, giveninits]{biblatex}
\usepackage{nicefrac,xfrac}

\usepackage{listings}
\usepackage{xcolor}
\usepackage{authblk}
\usepackage{bbm}

\newcommand{\N}{\mathbb{N}}
\newcommand{\Q}{\mathbb{Q}}
\newcommand{\R}{\mathbb{R}}
\newcommand{\Z}{\mathbb{Z}}
\newcommand{\e}{\mathrm{e}}
\newcommand{\E}{\mathbb{E}}

\title{Intrinsic Randomness in Epidemic Modelling Beyond Statistical Uncertainty}

\author[1.*]{Matthew J. Penn}
\author[2]{Daniel J.\ Laydon}
\author[1]{Joseph Penn}
\author[2]{Charles Whittaker}
\author[2]{Christian Morgenstern}
\author[2]{Oliver Ratmann}
\author[3]{Swapnil Mishra}
\author[4,2]{Mikko S. Pakkanen}
\author[1,2]{Christl A. Donnelly}
\author[2,5.*]{Samir Bhatt}

\affil[1]{University of Oxford, Oxford, UK}
\affil[2]{Imperial College London, London, UK}
\affil[3]{National University of Singapore, Singapore}
\affil[4]{University of Waterloo, Ontario, Canada}
\affil[5]{University of Copenhagen, Copenhagen, Denmark}
\affil[$^*$]{corresponding authors: matthew.penn@st-annes.ox.ac.uk, s.bhatt@imperial.ac.uk}

\date{}

\begin{document}

\maketitle
\begin{abstract}
 Uncertainty can be classified as either aleatoric (intrinsic randomness) or epistemic (imperfect knowledge of parameters). The majority of frameworks assessing infectious disease risk consider only epistemic uncertainty. We only ever observe a single epidemic, and therefore cannot empirically determine aleatoric uncertainty. Here, we characterise both epistemic and aleatoric uncertainty using a time-varying general branching process. Our framework explicitly decomposes aleatoric variance into mechanistic components, quantifying the contribution to uncertainty produced by each factor in the epidemic process, and how these contributions vary over time. The aleatoric variance of an outbreak is itself a renewal equation where past variance affects future variance. We find that, superspreading is not necessary for substantial uncertainty, and profound variation in outbreak size can occur even without overdispersion in the offspring distribution (i.e. the distribution of the number of secondary infections an infected person produces). Aleatoric forecasting uncertainty grows dynamically and rapidly, and so forecasting using only epistemic uncertainty is a significant underestimate. Therefore, failure to account for aleatoric uncertainty will ensure that policymakers are misled about the substantially higher true extent of potential risk. We demonstrate our method, and the extent to which potential risk is underestimated, using two historical examples.
  \end{abstract}

\subsection*{Introduction}

Infectious diseases remain a major cause of human mortality. Understanding their dynamics is essential for forecasting cases, hospitalisations, and deaths, and to estimate the impact of interventions. The sequence of infection events defines a particular epidemic trajectory -- the outbreak -- from which we infer aggregate, population-level quantities. The mathematical link between individual events and aggregate population behaviour is key to inference and forecasting. The two most common analytical frameworks for modelling aggregate data are susceptible-infected-recovered (SIR)  models \cite{Kermack1927-ow} or renewal equation models \cite{Fraser2007,Pakkanen2021-cc}. Under certain specific assumptions, these frameworks are deterministic and equivalent to each other \cite{Champredon2018-sg}. Several general stochastic analytical frameworks exist \cite{Allen2017-oo,Pakkanen2021-cc},  and to ensure analytical tractability make strong simplifying assumptions (e.g.\ Markov or Gaussian) regarding the probabilities of individual events that lead to emergent aggregate behaviour. 

We can classify uncertainty as either aleatoric (due to randomness) or epistemic (imprecise knowledge of parameters) \cite{Kiureghian2009-el}. The study of uncertainty in infectious disease modelling has a rich history in a range of disciplines, with many different facets \cite{Castro2020-vs,Neri2021-nc,scarpino2017}. These frameworks commonly propose two general mechanisms to drive the infectious process. The first is the infectiousness, which is a probability distribution for how likely an infected individual is to infect someone else. The second is the infectious period, i.e.\ how long a person remains infectious. The infectious period can also be used to represent isolation, where a person might still be infectious but no longer infects others and therefore is considered to have shortened their infectious period. Consider fitting a renewal equation to observed incidence data \cite{Pakkanen2021-cc}, where infectiousness is known but the rate of infection events $\rho(\cdot)$ must be fitted. The secondary infections produced by an infected individual will occur randomly over their infectious period $g$, depending on their infectiousness $\nu$. The population mean rate of infection events is given by $\rho(t)$, and we assume that this mean does not differ between individuals (although each individual has a different random draw of their number of secondary infections). In Bayesian settings, inference yields multiple posterior estimates for $\rho$, and therefore multiple incidence values. This is epistemic uncertainty: any given value of $\rho$ corresponds to a single realisation of incidence. However, each posterior estimate of $\rho$ is in fact only the mean of an underlying offspring distribution (i.e.\ the distribution of the number of secondary infections an infected person produces). If an epidemic governed by identical parameters were to happen again, but with different random draws of infection events, each realisation would be different, thus giving aleatoric uncertainty. 

When performing inference, infectious disease models tend to consider epistemic uncertainty only due to the difficulties in performing inference with aleatoric uncertainty (e.g.\ individual-based models) or analytical tractability. There are many exceptions such as the susceptible-infected-recovered model, which has stochastic variants that are capable of determining aleatoric uncertainty \cite{Allen2017-oo} and have been used in extensive applications (e.g.\ \cite{Pullano2021-dj}). However, we will show that this model can underestimate uncertainty under certain conditions. An empirical alternative is to characterise aleatoric uncertainty by the final epidemic size from multiple historical outbreaks \cite{Wong2020-yu,Cirillo2020-pt} but these are confounded by temporal, cultural, epidemiological, and biological context, and therefore parameters vary between each outbreak. Here, following previous approaches \cite{Allen2017-oo}, we analyse aleatoric uncertainty by studying an epidemiologically-motivated stochastic process, serving as a proxy for repeated realisations of an epidemic. Within our framework, we find that using epistemic uncertainty alone is a vast underestimate, and accounting for aleatoric uncertainty shows potential risk to be much higher.  We demonstrate our method using two historical examples: firstly the 2003 severe acute respiratory syndrome (SARS) outbreak in Hong Kong, and secondly the early 2020 UK COVID-19 epidemic.

\subsection*{Results}

\subsubsection*{An analytical framework for aleatoric uncertainty}

A time-varying general branching processes proceeds as follows: first, an individual is infected, and their infectious period is distributed with probability density function $g$ (with corresponding cumulative distribution function $G$). Second, while infectious, individuals randomly infect others (via a counting process with independent increments), driven by their infectiousness $\nu$ and a rate of infection events $\rho$. That is, an individual infected at time $l$, will, at some later time while still infectious $t$, generate secondary infections at a rate $\rho(t)\nu(t-l)$. $\rho(t)$ is a population-level parameter closely related to the time-varying reproduction number $R(t)$ (see Methods and  \cite{Pakkanen2021-cc} for further details), while $\nu(t-l)$ captures the individual's current infectiousness (note that $t-l$ is the time since infection).  We allow multiple infection events to occur simultaneously, and assume individuals behave independently once infected, thus allowing mathematical tractability \cite{Harris1963-sa}. Briefly, we model an individual's secondary infections using a stochastic counting process, which gives rise to secondary infections (i.e.\ offspring) that are either Poisson or Negative Binomial distributed in their number, and Poisson distributed in their timing (see Supplementary Notes 3.3 and 3.4). We study the aggregate of these events (prevalence or incidence) through closed-form probability generating functions and probability mass functions. Our approach models epidemic evolution through intuitive individual-level characteristics while retaining analytical tractability. Importantly, the mean of our process follows a renewal equation \cite{Parag2020-xe,Pakkanen2021-cc,EpiNow2}. Our formulation unifies mechanistic and individual-based modelling within a single analytical framework based on branching processes. Figure \ref{fig:Schematic} shows a schematic of this process. Formal derivation is in Supplementary Note 3. 

\begin{figure}[H]
\includegraphics[scale=1]{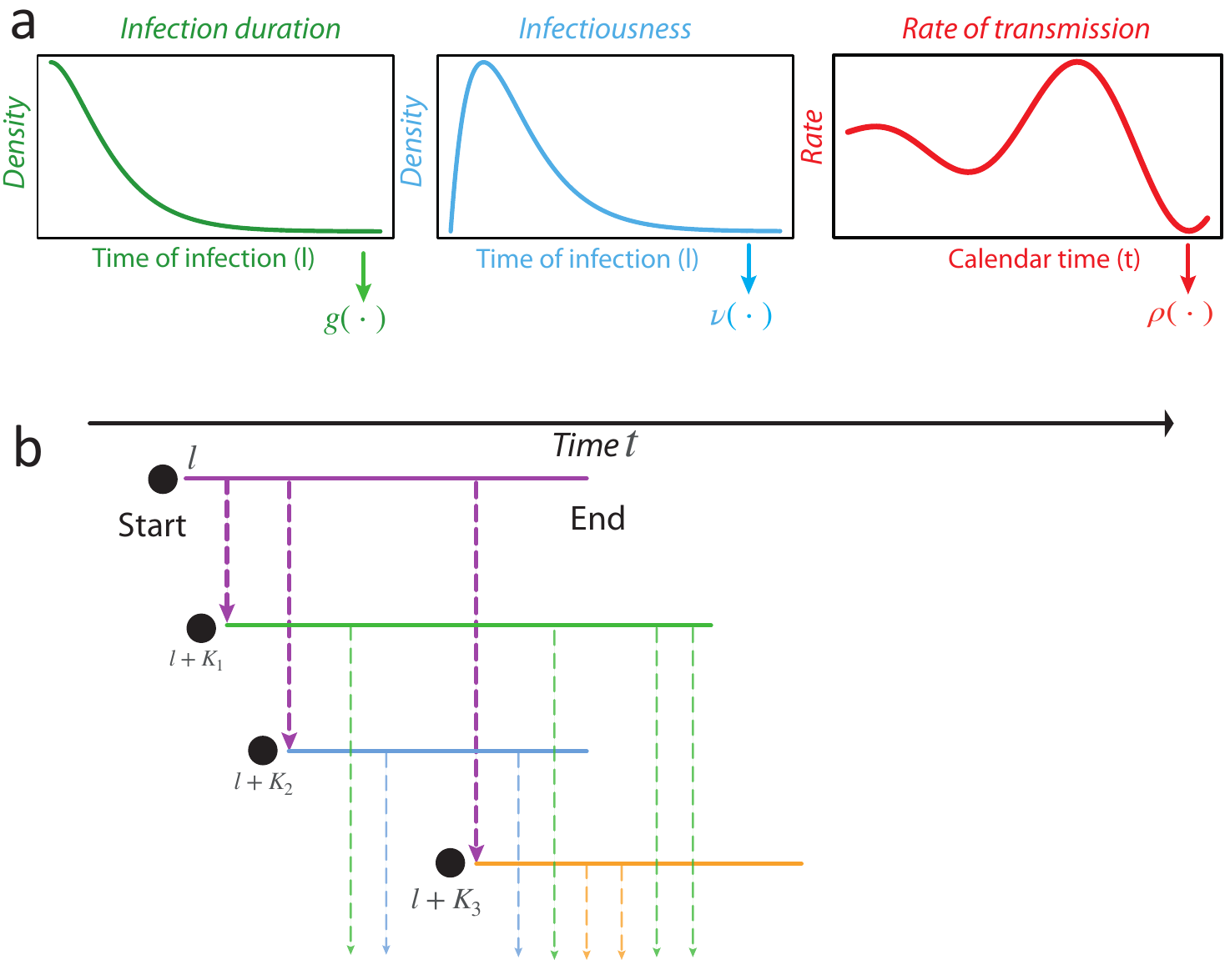}
\caption{Schematic of a time-varying general branching process. (a) shows schematics for the infectious period, an individual's time-varying infectiousness (both functions of time post infection $t^*$), and the population-level mean rate of infection events. The infectious period is given by probability density function $g$. For each individual their (time-varying) infectiousness and rate of infection events are given by $\nu$ and $\rho$ respectively. In an example (b), an individual is infected at time $l$, and infects three people (random variables $K$, purple dashed lines) at times $l+K_1$, $l+K_2$ and $l+K_3$. The times of these infections are given by a random variable with probability density function $\sim \frac{\rho(t)\nu(t-l)}{\int_l^{t}\rho(u)\nu(u-l) du}$. Each new infection then has its own infectious period and secondary infections (thinner coloured lines).}
\label{fig:Schematic}
\end{figure}

Randomness occurs at individual level, and there is a distribution of possible realisations of the epidemic given identical parameters. Simulating our general branching process would be cumbersome using the standard approach of Poisson thinning \cite{Ogata1981-gl}, and inference from simulation is more challenging still. Using probability generating functions, we analytically derive important quantities from the distribution of the number of infections, including the (central) moments and marginal probabilities given $\rho,g$ and $\nu$ (with or without epistemic uncertainty). We additionally use the probability generating function to prove general, closed-form, analytical results such as the decomposition of variance into mechanistic components, and the conditions under which overdispersion exists (i.e.\ where variance is greater than the mean). Finally, we derive a general probability mass function (likelihood function) for incidence. 

If infection event $k = 0, \ldots, n$ occurred at time $\tau_k$ and produced $y_k$ infections, let $x_{kj}$ denote the end time of the infectious period of the $j^{th}$ infection at event $k$. Note that $\tau_0 = l$ is the time of the first infection event and $y_0 = 1$. Then the likelihood $L_{\text{InfPeriod}}$ of each infected person's infectious period is a product over all infections given by \begin{equation}
L_{\text{InfPeriod}} = \prod_{k=0}^n \prod_{j=1}^{y_k} g(x_{kj} - \tau_k,\tau_k).
\end{equation} The likelihood of there being $y_k$ infections at time $\tau_k$ is given by \begin{equation}
L_{\text{InfTime}} = \prod_{k=1}^n \bigg( \sum_{i = 0}^{k-1}\sum_{j=1}^{y_i} \mathds{1}_{\{x_{ij} < \tau_k\}} p_{y_k}(\tau_k,\tau_i) \bigg),
\end{equation} where $p_{y_k}(\tau_k,\tau_i)$ is the (infinitesimal) rate at which an individual infected at $\tau_i$ causes $y_k$ infections at time $\tau_k$, provided it is still infectious. Finally, the probability that no other infections occurred between the infection events at times $(\tau_k)_{k=0}^n$ is given by \begin{equation}
L_{\text{Only}} = \exp\bigg(-\sum_{i=0}^n\sum_{j=1}^{y_i}\int_{\tau_i}^{\min(t,x_{ij})}r(u,\tau_i)du\bigg),
\end{equation} where $r$ is the infection event rate and $t$ is the current time. Note the term $\text{exp}(-x)$ comes from a Poisson assumption.  Our full likelihood $L_{\text{Full}}$ is then \begin{equation}
L_{\text{Full}} = L_{\text{InfPeriod}} \times L_{\text{InfTime}} \times L_{\text{Only}}.
\label{eq:likelihood} 
\end{equation}
Full derivations of these quantities are provided in Supplementary Note 3. If discrete time is assumed, equation \ref{eq:likelihood} simplifies to a likelihood commonly used for inference \cite{Cori2013}. Markov Chain Monte Carlo can be used on equation \ref{eq:likelihood} to sample aleatoric incidence realisations, but it is often simpler to solve the probability generating function with complex integration. The probability generating function, equations for the variance, and derivations of the probability mass function are found in Supplementary Notes 3,4,5 and 6, and a summary of the main analytical results is found in the Methods. 

\subsubsection*{The dynamics of Uncertainty}

We derive the mean and variance of our branching process. The general variance Equation \ref{eq:variance_renewal} (see Methods) captures uncertainty in  prevalence over time, where individual-level parameters govern each infection event. This equation comprises three terms: the timing of secondary infections from the infectious period (Equation \ref{eq:variance_renewal}a); the offspring distribution (Equation \ref{eq:variance_renewal}b); and propagation of uncertainty through the descendants of the initial individual (Equation \ref{eq:variance_renewal}c). Importantly, this last term depends on past variance, showing that the infection process itself contributes to aleatoric variance, and does not arise only from uncertainty in individual-level events. In short, unlike common Gaussian stochastic processes, the general variance in disease prevalence is described through a renewal equation. Therefore, future uncertainty depends on past uncertainty, and so the uncertainty around subsequent epidemic waves has memory. Additionally, uncertainty is driven by a complex interplay of time-varying factors, and not simply proportional to the mean. For example, a large first wave of infection can increase the variance of the second wave. As such, the general variance equation \ref{eq:variance_renewal} disentangles and quantifies the causes of uncertainty, which remain obscured in brute-force simulation experiments \cite{Allen2017-oo}. 

\begin{figure}[H]
\includegraphics[scale=0.47]{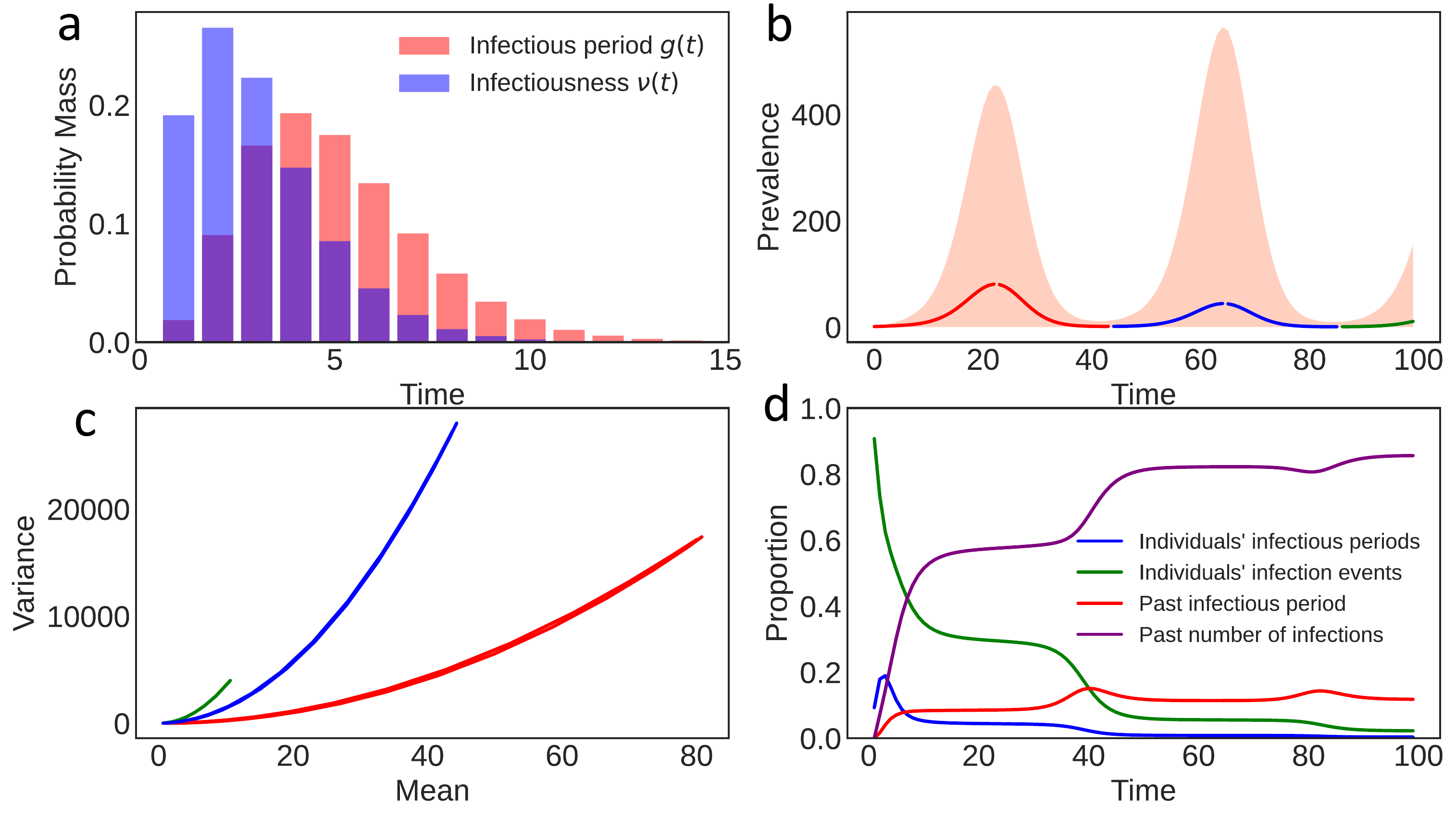}
\caption{Aleatoric uncertainty without overdispersed offspring distribution. Plots show simulated epidemic where $\rho(t) = 1.4 + \text{sin}(0.15t)$, with a Poisson offspring distribution. We use infectiousness $\nu \sim \text{Gamma}(3,1)$, and infectious period $g \sim \text{Gamma(5,1)}$. (a) Overlap between $g$ and the infectiousness $\nu$, where $g$ controls when the infection ends e.g.\ by isolation. (b) Predicted mean and 95\% aleatoric uncertainty intervals for prevalence. Note there is no epistemic uncertainty as the parameters are known exactly (c) Phase plane plot showing the mean plotting against the variance. (d) Proportional contribution to the variance from the individual terms in Equation \ref{eq:variance_renewal}. Compounding uncertainty from past events is the dominant contributor to overall uncertainty.}
\label{fig:Figure0}
\end{figure}

Consider a toy simulated epidemic with $\rho(t) = 1.4 + \text{sin}(0.15t)$, where the offspring distribution is Poisson in both timing and number of secondary infections, and where infectiousness $\nu$ is given by the probability density function $\nu\sim \text{Gamma}(3,1)$, and, similarly, the infectious period $g \sim \text{Gamma(5,1)}$. Here the parameters of the Gamma distribution are the shape and scale respectively. The resulting variance is counterintuitive. We prove analytically that overdispersion emerges despite a non-overdispersed Poisson offspring distribution. The second wave has a lower mean but a higher variance than the first wave (Figure \ref{fig:Figure0}), because uncertainty is propagated. If the variance were Poisson, i.e.\ equal to the mean, the second wave would instead have a smaller variance due to fewer infections. Initially, uncertainty from individuals is largest, but as the epidemic progresses, compounding uncertainty propagated from the past dominates [Figure \ref{fig:Figure0}, bottom right]. Note that in this example with zero epistemic uncertainty (we know the parameters perfectly), aleatoric uncertainty is large.

In Equation \ref{eq:variance_renewal}, the first two terms account for uncertainty in the infectious periods of all infected individuals. The third term denotes the uncertainty from the offspring distribution. By construction, the timing of infections is an inhomogenous Poisson process, where at each infection time the number of infections is random. The third term (Equation \ref{eq:variance_renewal}b) contains the second moment of the offspring distribution, which is the variability around its mean (i.e.\ $\rho(t)$). The second moment quantifies the extent of possible superspreading. In contrast to other studies \cite{Woolhouse1997-vn,Lloyd-Smith2005-ge}, we find that individual-level overdispersion in the offspring distribution is less important than explosive epidemics. Under a null Poisson model, with no overdispersion (see Poisson case in Figure \ref{fig:Figure0}), substantial aleatoric uncertainty arises from a Poisson offspring distribution combined with variance propagation. We rigorously prove via the Cauchy-Schwarz inequality that, under a mild condition on the possible spread of the epidemic, the variance of number of infections at a given time is always greater than the mean, and hence is overdispersed. Overdispersion in the offspring infection distribution is therefore not necessary  for high aleatoric uncertainty, although it still increases variance at both individual-level and population-level. 

We derive the conditional variance, with known past events but unknown future events. Conditional variance grows proportionally to the square of the mean, with  additional terms containing the previous variance. Therefore aleatoric uncertainty grows and forecasting exercises based only on epistemic uncertainty greatly underestimates the risk of very large epidemics, and this underestimation becomes more severe as the forecast horizon expands or as the epidemic grows. 

\subsubsection*{Aleatoric uncertainty over the SARS 2003 epidemic}

\begin{figure}[H]
\centering
\includegraphics[scale=0.34,left]{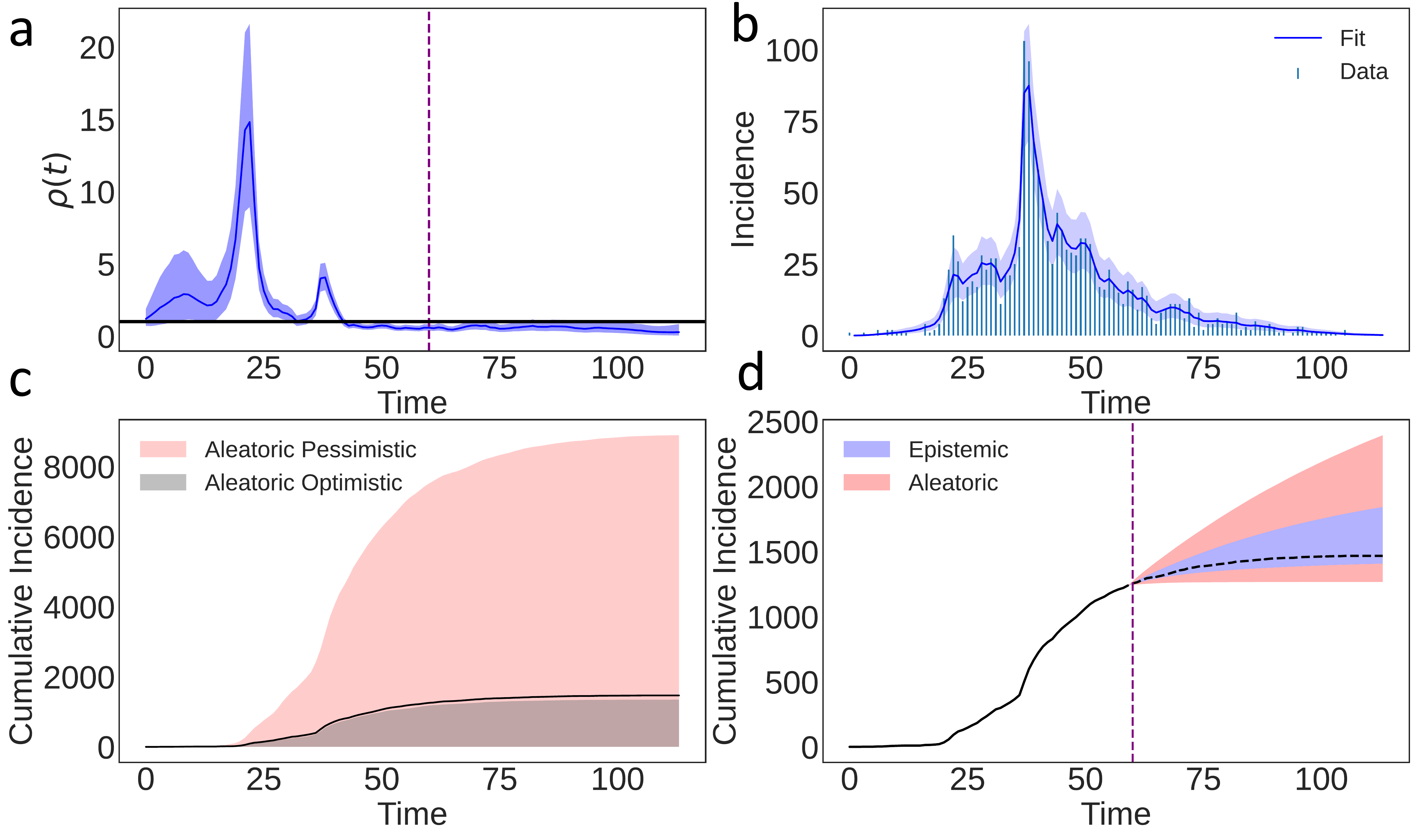}
\caption{The 2003 SARS epidemic in Hong Kong \cite{Lipsitch2003-gy,Cori2013-jc}. (a) $\rho(t)$ with 95\% epistemic uncertainty. (b) Fitted incidence mean, 95\% epistemic uncertainty with observational noise from using Equation \ref{eq:likelihood}. Data is daily incidence of symptom onset. (c) Aleatoric uncertainty from the start of the epidemic under an optimistic and pessimistic $\rho(t)$. (d) Epistemic (blue) and epistemic and aleatoric uncertainty (red) while keeping $\rho$ constant at the forecast data (dotted line). Forecasting is from day 60.
}
\label{fig:SARS}
\end{figure}

To demonstrate the importance of aleatoric uncertainty, we analyse daily incidence of symptom onset in Hong Kong during the 2003 severe acute respiratory syndrome (SARS) outbreak \cite{Lipsitch2003-gy,Cori2013-jc,Hung2003-wu}. The epidemic struck Hong Kong in March-May 2003, with a case fatality ratio of 15\%. We fit a Bayesian renewal equation assuming a random walk prior distribution for the rate of infection events $\rho$ \cite{Pakkanen2021-cc}, using Equation \ref{eq:likelihood} for inference. We ignore $g$ and assume that the distribution of generation times mirrors the distribution of infectiousness, i.e.\ that the infectiousness $\nu$ equals the generation time \cite{Lipsitch2003-gy}. Note these parameter choices are illustrative and do not affect our main conclusions. The fitted $\rho(t)$ in Figure \ref{fig:SARS} (top left) shows two major peaks, consistent with the major transmission events in the epidemic \cite{Hung2003-wu}. Figure \ref{fig:SARS} (top right) shows the mean epistemic fit, with epistemic (posterior) uncertainty tightly distributed around the data. Figure \ref{fig:SARS} (bottom left) shows the aleatoric uncertainty under optimistic and pessimistic scenarios (i.e.\ the upper and lower bounds of $\rho(t)$ in Figure \ref{fig:SARS} (top right)). The pessimistic scenario includes the possibility of extinction, but also an epidemic that could have been more than six times larger than that observed. The optimistic scenario suggests we would observe an epidemic of at worst comparable size to that observed. Finally, Figure \ref{fig:SARS} (bottom right) shows epistemic and aleatoric forecasts at day 60 of the epidemic, fixing $\rho(t)$ using the 95\% epistemic uncertainty interval to be constant at either $\rho(t\geq 60)=0.38$ or $\rho(t\geq60) = 0.83$ and simulating forwards. While the epistemic forecast does contain the true unobserved outcome of the epidemic, it underestimates true forecast uncertainty, which is 1.3 times larger. The range of the constant $\rho$ for forecast is below 1, and yet we still see substantial aleatoric uncertainty. If $\rho$ were above 1 for a sustained period, aleatoric uncertainty would play a smaller role \cite{Barbour2013-xo}, but this is rare with real epidemics, where susceptible depletion, behavioural changes or interventions keep $\rho$ around 1. Our results therefore highlight that epistemic uncertainty drastically underestimates potential epidemic risk.

\subsubsection*{Aleatoric risk assessment in the early 2020 COVID-19 pandemic in the UK}

To demonstrate the practical application of our model, we retrospectively examine the early stage of the COVID-19 pandemic in the UK, using only information available at the time. While the date of the first locally transmitted case in the UK remains unknown (likely mid-January 2020 \cite{Pybus2020-nu}), COVID-19 community transmission was confirmed in the UK by late January 2020, and we therefore start our simulated epidemic on January 31st 2020. We consider uncertainty in the predicted number of deaths on March 16th 2020 \cite{Ferguson2020-yn}, during which time decisions regarding non-pharmaceutical interventions were made. Testing was extremely limited during this period, and COVID-19 death data were unreliable. For this illustration, we assume that we did not know the true number of COVID-19 deaths, as was the case for many countries in early 2020. Policymakers then needed estimates of the potential death toll, given limited knowledge of COVID-19 epidemiology and unreliable national surveillance. 

\begin{figure}
\centering
\includegraphics[scale=0.4,left]{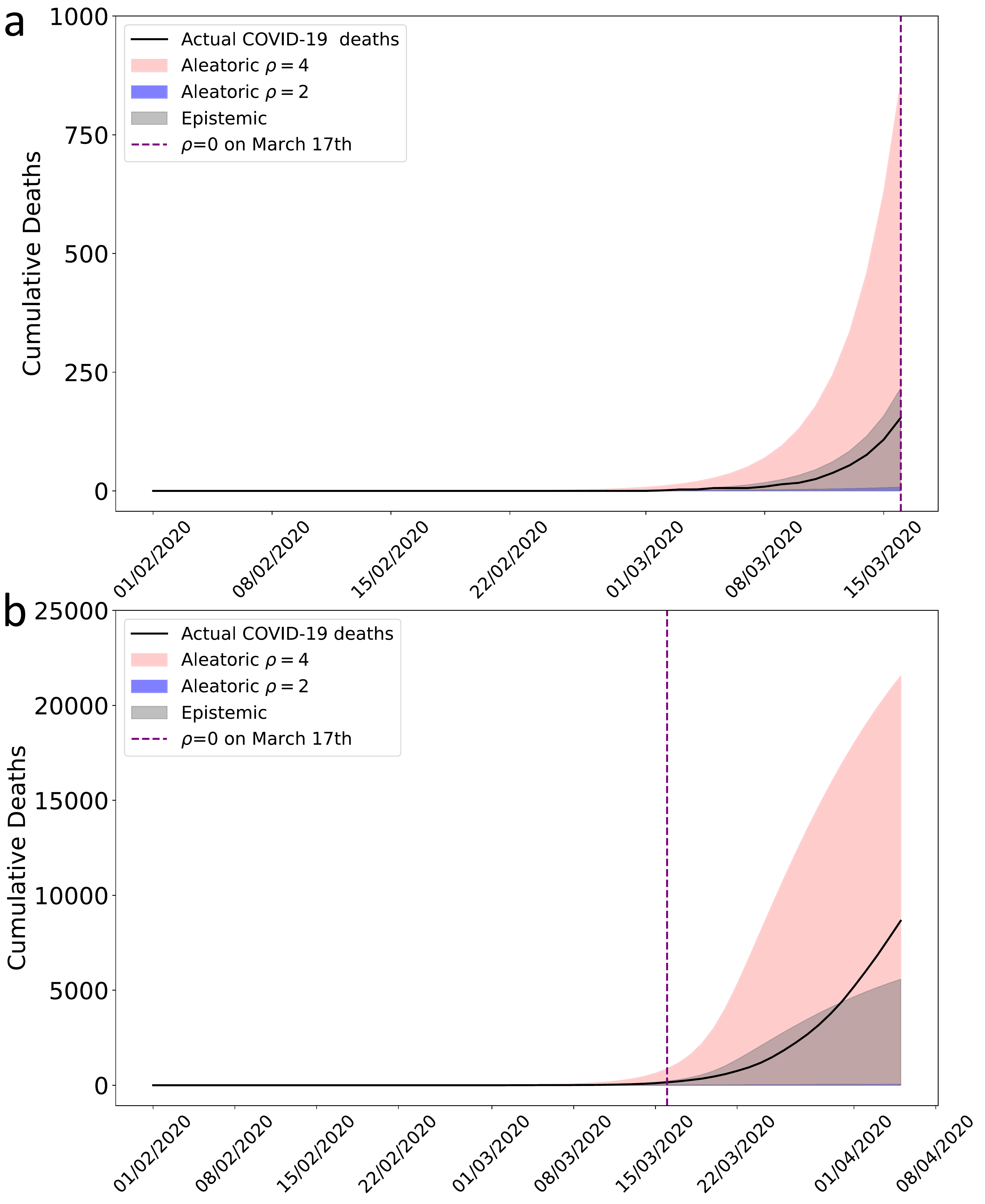}
\caption{Early 2020 COVID-19 pandemic in the UK. (a) shows a simulated epidemic using parameters available on March 16th 2020 (Table \ref{tab:table1}), for a plausible range of $\rho=R_0$ between 2 and 4. Blue bars indicate actual COVID-19 deaths, which we assume no knowledge of. The purple line is March 17th 2020, we set transmission to zero i.e.\ $\rho=0$, to simulate an intervention that stops transmission completely. The grey envelope is the epistemic uncertainty and the red envelope the aleatoric uncertainty. (b) is the same as the top plot, except time is extended past March 17th with transmission being zero.  Note aleatoric uncertainty is presented but is very close to zero.
}
\label{fig:Figure2}
\end{figure}

We simulated an epidemic from a time-varying general branching process with a  Negative Binomial offspring distribution, using parameters that were largely known by March 16th 2020 (Table \ref{tab:table1}). The infection fatality ratio, infection-to-onset distribution and onset-to-death distribution were convoluted with incidence \cite{Pakkanen2021-cc} to estimate numbers of deaths. Estimated COVID-19 deaths and uncertainty estimates between January 31st and March 16th 2020 are shown in Figure \ref{fig:Figure2} (Top). While the epistemic uncertainty contains the true number of deaths, it is still an underestimate, and including aleatoric uncertainty, we find that the epidemic could have had more than four times as many deaths. Consider a hypothetical intervention on March 17th 2020 (Figure \ref{fig:Figure2} (bottom)) that completely stops transmission. Deaths would still occur from those already infected but no new infections would arise. In this hypothetical case, the aleatoric uncertainty would still be 2.5 times the actual deaths that occurred (when in fact transmission was never zero or close to it). This hypothetical scenario highlights the scale of aleatoric uncertainty, and demonstrates that our method can be useful in assessing risk in the absence of data by giving a reasonable worst case. Further, we observe that using only epistemic uncertainty provides a reasonably good fit in a relatively short time-horizon (Figure \ref{fig:Figure2}, Top), but soon afterwards greatly underestimates uncertainty (Figure \ref{fig:Figure2}, Bottom). The fits using aleatoric uncertainty provide a more reasonable assessment of uncertainty. While we concentrate on the upper bound, the lower bound on the worst-case scenario still exceeds zero, and therefore the epidemic going extinct by March 16th in the worst-case with no external seeding would have been very unlikely. Aleatoric uncertainty highlights a more informative reasonable worst-case estimate than epistemic uncertainty alone, and could be a useful metric for a policymaker in real time, with low-quality data, without requiring simulations from costly, individual-based models.
\begin{table}[]
\centering
\begin{tabular}{|l|l|l|}
\hline
\textbf{Epidemiological Parameter} & \textbf{Value or Distribution}           & \textbf{Citation}                                                            \\ \hline
Infection Fatality Ratio            & 0.9\%                                    & \cite{verity_estimates_2020,Brazeau2022-ik}               \\ \hline
Basic Reproduction Number          & $2-4$                                    & \cite{Ferguson2020-yn,Liu2020-ut}                           \\ \hline
Serial Interval Distribution       & $\sim \text{Gamma}(7.82,\frac{1}{0.62})$ & \cite{verity_estimates_2020,Flaxman2020-zg,Sharma2021-cs} \\ \hline
Onset-to-Death Distribution        & $\sim \text{Gamma}(1.45, 10.43)$         & \cite{verity_estimates_2020,Mishra2020-yt}                  \\ \hline
Infection-to-Onset Distribution    & $\sim \text{Gamma}(35.16, 6.9)$          & \cite{Flaxman2020-zg,verity_estimates_2020}               \\ \hline
Overdispersion Coefficient         & 0.53                                     & \cite{kucharski_early_2020}                               \\ \hline
\end{tabular}
\caption{\label{tab:table1} Epidemiological parameters available on March 16th 2020 used in branching process simulation}
\end{table}

\subsection*{Discussion}

Stochastic models more realistically model natural phenomena than deterministic equations \cite{Mumford_undated-ri}, and particularly so with infection processes \cite{Anderson1972-uh}. Accordingly, individual-based models have found much success \cite{Willem2017-hy,Ferguson2005-nv} in capturing the complex dynamics that emerge from infectious disease outbreaks, and have been highly influential in policy \cite{Ferguson2020-yn}. However, despite a plethora of alternatives, many analytical frameworks still tend to be deterministic \cite{Cori2013-jc,Flaxman2020-zg,Faria2021-sp}, and only consider statistical, epistemic parameter uncertainty. Frameworks that expand deterministic, mechanistic equations to include stochasticity use a Gaussian noise process \cite{Allen2017-oo}, or restrict the process to be Markovian. Markovian branching processes require the infection period or generation time to be exponentially distributed - a fundamentally unrealistic choice for most infectious diseases. Further, a Gaussian noise process is unlikely to be realistic \cite{Cirillo2020-pt}.

Our results show that individual-level uncertainty is overshadowed by uncertainty in the infection process itself. Profound overdispersion in infectious disease epidemics is not simply a result of overdispersion in the offspring distribution, but is fundamental and inherent to the branching process. We rigorously prove that even with a Poisson offspring distribution (not characterized by overdispersion), overdispersion in resulting prevalence or incidence is still virtually always guaranteed. We show that forecast uncertainty increases rapidly, and therefore common forecasting methods almost certainly underestimate true uncertainty. Similar to other existing frameworks, our approach provides a different methodological tool to evaluate uncertainty in the presence of little to no data, assess uncertainty in forecasting, and  retrospectively assess an epidemic. Other approaches, such as agent based models, could also be readily used. However, the framework we present permits the unpicking of dynamics analytically and from first principles without a black box simulator. Equally, this is also a limitation, since new and flexible mechanisms cannot be easily integrated or considered. 

We have considered only a small number of mechanisms that generate uncertainty. Cultural, behavioural and socioeconomic factors could introduce even greater randomness. Therefore our framework may underestimate true uncertainty in infectious disease epidemics. The converse is also likely, contact network patterns and spatial heterogeneity also limit the routes of transmission, such that the variability in anything but a fully connected network will be lower. Furthermore, our assumption of homogeneous mixing and spatial independence overestimates uncertainty. A sensible next step for future research to to study the dynamics of these branching processes over complex networks. Finally at the core of all branching frameworks in an assumption of independence, which is unlikely to be completely valid (people mimic other people in their behaviour) but is necessary for analytical tractability. Studying the effect of this assumption compared to agent based models would also be a useful area of future research.

We provide one approach to determining aleatoric uncertainty. Other approaches based on stochastic differential equations, Markov processes, reaction kinetics, or Hawkes processes all have their respective advantages and disadvantages. The differences in model specific aleatoric uncertainty and how close the models come to capturing the true, unknown, aleatoric uncertainty is a fundamental question moving forwards. In this paper we have provided yet another approach to characterise aleatoric uncertainty, where this approach is most useful and how it can be reconciled with existing approaches will be an interesting area of study.

\subsection*{Methods}

Detailed derivations of the methods can be found in the Supplementary Notes, with a high level description of the content found in Supplementary Note 1.

A time-varying general branching process proceeds as follows: first, a single individual is infected at some time $l$, and their infectious period $L$ is distributed with probability density function $g$ (and cumulative distribution function $G$). Second, during their infectious period, they randomly infect other individuals, affected by their infectiousness $\nu(t-l)$, and their mean number of secondary infections, which is assumed to be equal to the population-level rate of infection events $\rho(t)$. $\rho(t)$ is closely related to the time-varying reproduction number $R(t)$ (see \cite{Pakkanen2021-cc} for details). The infectious period $g$ accounts for variation in individual behaviour. If people take preventative action to reduce onward infections, their reduced infection period can stop transmission despite remaining infectious. Where infectious individuals do not change their behaviour, $g$ can be ignored and individual-level transmission is controlled by infectiousness $\nu$ only. Each newly infected individual then proceeds independently by the same mechanism as above.  Specifics can be found in Supplementary Notes 2.1-2.5.

Formally, if an individual is infected at time $s$, their number of secondary infections is given by a stochastic counting process $\{N(t,s)\}_{t\geq s}$, which is independent of other individuals and has independent increments. We assumehere that the epidemic occurs in continuous time, and hence that $N(t,s)$ is continuous in probability, although we consider discrete-time epidemics in Supplementary Note 7. To aid calculation, we suppose $N(t,s)$ can be defined from a Lévy Process $\mathcal{N}(t)$ - that is, a process with both independent and identically distributed increments - via $N(t,s) = \mathcal{N}\bigg(\int_s^tr(k,s)dk\bigg)$ for some non-negative rate function $r$. It is assumed that each counting process $\{N(t,s)\}_{t\geq s}$ is defined from an independent copy of $M(t)$. This formulation has two advantages: first, the dependence of $N(t,s)$ on $s$ is restricted to the rate function $r$; and second, if $J_{\mathcal{N}}(t)$ counts the number of infection events in $\mathcal{N}(t)$ (where here infection events refer to an increase, of any size, in $N(t,s)$), then $J_{\mathcal{N}}(t)$ is a Poisson process with some rate $\kappa$ \cite{Applebaum2009-km}. We can then define $J(t,s)$ to be the counting process of infection events in $N(t,s)$, and $Y(v)$ to be size of the infection event (i.e.\ the number of secondary infections that occur) t time $v$. We assume that $Y$ is independent of $s$, although such a dependence would curtail superspreading to depend on infectiousness, and could be incorporated into the framework. Therefore $J(t,s)$ is an inhomogeneous Poisson Process (and so $N(t,s)$ has been characterised as an inhomogeneous compound Poisson Process). We consider the cases where $N(t,s)$ is itself an inhomogeneous Poisson process, and where $N(t,s)$ is a Negative Binomial process. This allows us to examine effects of overdispersion in the number of secondary infections, although our framework allows for more complicated distributions.

Here, $r(t,l) = \rho(t)\nu(t-l)$ where $\rho(t)$ models the population-level rate of infection events, and $\nu(t-l)$ models the infectiousness of an individual infected at time $l$. If $\nu(t-l)$ is sufficiently well characterised by the generation time (i.e. where the timing of secondary infections mirrors tracks their infectiousness) , and the infectious period can be ignored, then the integral $\int_l^t r(s,l) ds$ has the same scale as the commonly used reproduction number $R(t)$ \cite{Pakkanen2021-cc}. The branching process yields a series of birth and death times for each individual (i.e.\ the time of infection and the end of the infectious period respectively), from which prevalence (the number of infections at any given time) or cumulative incidence (the total number of infections up to any time) can be defined.

\subsubsection*{Probability generating function}
We derive the probability generating function for a time-varying age-dependent branching process, allowing derivation of the mean and higher-order moments (full derivations can be found in Supplementary Notes 3.1-3.7). We consider two special cases for the number of new infections $Y(v)$ at each infection event: a Poisson distribution and a logarithmic (log series) distribution. In both cases, we assume that the distribution of $Y(v)$ is equal for all values of $v$. In the Poisson case, the number of new infections at each infection time is, by definition, one. Therefore the number of infections an individual creates is Poisson distributed, and closely clustered around the mean rate of infection events. The logarithmic case, which causes $N(t,l)$ to be a Negative Binomial process, more realistically allows multiple infections to occur at each infection time, and so the number of infections an individual causes is overdispersed. The pgf (probability generating function), $F(t,l;s) = E(s^{Z(t,l)})$, can be derived by conditioning on the lifetime, $L$, of the first individual. That is,
\begin{equation}
    E\bigg(s^{Z(t,l)}\bigg)  = \bigg(1-G(t-l,l)\bigg) E\bigg(s^{Z(t,l)}\bigg|L \geq t-l\bigg) + \int_0^{t-l}E\bigg(s^{Z(t,l)}\bigg|L = u\bigg) g(u,l)du
\end{equation}
Note that if the individuals directly infected by the initial individual are infected at times $l+t_1,...,l+t_n$, then
\begin{equation}
    Z(t,l) = 1 + \sum_{i=1}^nZ(t,l+t_i)
\end{equation}
This observation allows us to write the generating function $F(t,l)$ as a function of $F(t,u)$ for $u \in (t,l)$. As $F(t,t) = s$, this allows us to iteratively find the value of $F(t,l)$. Explicitly, we have
\begin{align}
\label{eq:pgf}
\underbrace{F(t,l;s)}_{\text{pgf}} &= \underbrace{\left(1-G(t-l,l)\right)}_{P(L > t-l)}\underbrace{q_1\left(\int_0^{t-l}\overbrace{f(\underbrace{ F(t,l+k;s)}_{\text{pgf of process started at $l+k$}})}^{\text{pgf of $Y$}}\overbrace{\rho(l+k)\nu(k)dk}^\text{infection rate at time $l+k$}  \right)}_{\text{pgf of $J(t,l)$}} \nonumber\\
\nonumber \\
&+\int_0^{t-l}\underbrace{q_2\left( \int_0^u \overbrace{f(\underbrace{ F(t,l+k;s)}_{\text{pgf of process started at $l+k$}})}^{\text{pgf of $Y$}} \overbrace{\rho(l+k)\nu(k)}^{\text{infection rate at time $l+k$}}dk \right)}_{\text{pgf of $J(l+u,l)$}}\underbrace{g(u,l) du}_{P(L = l+u)},
\end{align}
where $q_1(z;s) = se^z$, and where $q_2(z)=e^z$ in the case where $Z(t,l)$ refers to prevalence, whereas $q_2(z;s)=se^z$ in the case where $Z(t,l)$ refers to cumulative incidence. Note also that $f(z) = z-1$ in the Poisson case and $f(z) = -\phi\left[\log\left(1-(1  -\frac{\phi}{1+\phi}z)\right) - \log\left(\frac{\phi}{1+\phi}\right)\right] $ in the log-series case and that the constant $\kappa$ is absorbed into $\rho$.

The key intuition in understanding Equation \ref{eq:pgf} is that for an integer random variable $X$ and iid (independent and identically distributed) random variables $Y_i$, $E(s^{\sum_{i=1}^X Y_i}) = G_X(G_Y(s))$, where $G_X$ and $G_Y$ are the generating functions of $X$ and $Y_i$ respectively. Thus, we expect the pgfs of the various parts of our model to combine via composition, as occurs in the equation above.

Mean incidence can recovered from both prevalence (via back calculation \cite{Pakkanen2021-cc}) and cumulative incidence. In Equation \ref{eq:pgf} for the Negative Binomial case, $\phi$ is the degree of overdispersion. Equation \ref{eq:pgf} is solvable using via quadrature and the fast Fourier transform via a result from complex analysis \cite{Lyness1967-yk} and scales easily to populations with millions of infected individuals, and the probability mass function can be computed to machine precision (a full derivation is available in Supplementary Note 3.7).

\subsubsection*{Variance decomposition}

For simplicity, we only summarise the decomposition for prevalence, but an analogous and highly similar derivation for cumulative incidence can be found in Supplementary Note 3.5.  We can derive an analytical equation for the mean and variance of the entire branching process (full derivations can be found in Supplementary Notes 4.1-4.7 and the mathematical properties of the variance equations can be found in Supplementary Notes 6.1-6.3) . 
The mean prevalence $M(t,l)$ is given by \begin{equation}
    M(t,l) = (1-G(t-l,l)) + \int_0^{t-l} M(t,l+u)\rho(l+u)\nu(u)\mathbb{E}(Y) (1-G(u,l))du.
  \label{eq:mean_renewal}
\end{equation}
Note, $\rho$ can be scaled to absorb the $E(Y)$ and $\kappa$ constants. Equation \ref{eq:mean_renewal} is consistent with that previously derived in \cite{Pakkanen2021-cc}. The second moment, $W(t,l) := \mathbb{E}(Z(t,l)(Z(t,l) - 1))$ allows us to determine the variance, $V(t,l)$ as $V(t,l) = W(t,l) + M(t,l) - M(t,l)^2$. 
The variance can be decomposed into three mechanistic components. 
\begin{align}
    V(t,l) &= \underbrace{\int_0^{t-l} \left[\int_0^u M(t,l+k)\rho(l+k)\nu(k) dk\right]^2 g(u,l)du - M(t,l)^2 }_{\text{(9a): uncertainty from the infectious period
    }}\nonumber\\    
    &+\underbrace{(1-G(t-l,l))\bigg[1 + 2\int_0^{t-l} M(t,l+u)\rho(l+u)\nu(u)du +  \bigg(\int_0^{t-l}M(t,l+u)\rho(l+u)\nu(u)du \bigg)^2\bigg]}_{\text{(9a continued): uncertainty from the infectious period}}\nonumber\\ 
      &+ \underbrace{\int_0^{t-l}M(t,l+u)^2\mathbb{E}(Y^2)\rho(l+u)\nu(u)(1-G(u,l))du}_{\text{(9b): uncertainty from the offspring distribution}}\nonumber\\
     &+\underbrace{\int_0^{t-l}V(t,l+u)\rho(l+u)\nu(u)(1-G(u,l))du}_{\text{(9c): uncertainty propagated from the past}}. 
     \label{eq:variance_renewal}
\end{align}
The general variance equation \ref{eq:variance_renewal} captures the evolution of uncertainty in population-level disease prevalence over time, where fixed individual-level disease transmission parameters govern each infection event. Unlike the simple Galton-Watson process, we find that previously unknown factors also determine aleatoric variation in disease prevalence. Specifically, the general variance equation \ref{eq:variance_renewal} comprises three terms, one for the infectious period (Equation \ref{eq:variance_renewal}a), one for the number and timing of secondary infections (Equation \ref{eq:variance_renewal}b), and a term that propagates uncertainty through descendants of the initial individual (Equation \ref{eq:variance_renewal}c). Importantly, the last term (Equation \ref{eq:variance_renewal}c) depends on past variance, showing that the infection process itself contributes to aleatoric variance, and this is distinct from the uncertainty in individual infection events. In short, and unlike Gaussian stochastic processes, the general variance in disease prevalence is described through a renewal equation. Intuitively then, uncertainty in an epidemic's future trajectory is contingent on past infections, and that the uncertainty around consecutive epidemic waves are connected.  As such, the general variance equation \ref{eq:variance_renewal} allows us to disentangle important aspects of infection dynamics that remain obscured in brute-force simulations \cite{Allen2017-oo}. 

\subsubsection*{Overdispersion}
We define an epidemic to be expanded if at time $t$ there is a non-zero probability that the prevalence, not counting the initial individual or its secondary infections, is non-zero.

Note that this is a very mild condition on an epidemic - in a realistic setting, the only way for an epidemic to not be expanded is if it is definitely extinct by time $t$, or if $t$ is small enough that tertiary infections have not yet occurred.

Large aleatoric variance intrinsic to our branching process implies that the prevalence of new infections (that is, prevalence excluding the deterministic initial case) is always strictly overdispersed at time $t$, providing the epidemic is expanded at time $t$. A full proof is given in Supplementary Note 4.4, but we provide here a simpler justification in the special case that $G(t-l,l) = 1$.

In this case,  prevalence of new infections is equal to  standard prevalence, and the equations for $M(t,l)$ and $V(t,l)$ simplify significantly. Switching the order of integration in the equation for $M(t,l)$ gives
\begin{equation}
 M(t,l) =  \int_0^{t-l} M(t,l+u)\rho(l+u)\nu(u) \mathbb{E}(Y)(1-G(u,l))du =  \int_0^{t-l}\bigg[\int_0^u M(t,l+k)\rho(l+k)\mathbb{E}(Y)\nu(k)dk\bigg]g(u,l)du 
\end{equation}
and hence, the Cauchy-Schwarz Inequality shows that
\begin{equation}
\label{eq:CS}
    M(t,l)^2 \leq \bigg(\int_0^{t-l}\bigg[\int_0^u M(t,l+k)\mathbb{E}(Y)\rho(l+k)\nu(k)dk\bigg]^2g(u,l)du\bigg)
\end{equation}
as $\int_0^{t-l}g(u,l)du = 1$. Thus, the first term, (\ref{eq:variance_renewal}a), in the variance equation is non-negative. 

The remaining terms can be dealt with as follows. (\ref{eq:variance_renewal}a) is equal to zero, and the sum of (\ref{eq:variance_renewal}c) is (using $Y(l+u,l)^2 \geq Y(l+u,l)$) bounded below by $\int_0^{t-l} \mathbb{E}(Z(t,l+u)^2)\mathbb{E}(Y)\rho(l+u)\nu(u)(1-G(u,l)) du$. Finally, noting that $Z(t,l+u)^2 \geq Z(t,l+u)$, this is bounded below by $\int_0^{t-l} M(t,l+u)\mathbb{E}(Y)\rho(l+u)\nu(u)(1-G(u,l))du = M(t,l)$. Hence, $V(t,l) \geq M(t,l)$ holds.

To show strict overdispersion, note that for $V(t,l) = M(t,l)$ to hold, it is necessary that
\begin{equation}
      \int_0^{t-l} \mathbb{E}(Z(t,l+u)^2)\mathbb{E}(Y)\rho(l+u)\nu(u) (1-G(u,l))du = \int_0^{t-l} M(t,l+u)\mathbb{E}(Y)\rho(l+u)\nu(u) (1-G(u,l))du
\end{equation}
and hence, for each $u$ (as $\mathbb{E}(Y) > 0$)
\begin{equation}
\label{eq:condition}
    \mathbb{E}\bigg[Z(t,l+u)(Z(t,l+u)-1)\bigg] = 0 \quad \text{or} \quad \rho(l+u)\nu(u) (1-G(u,l))=0
\end{equation}
If new infections can be caused, then more than one new infection can be caused. Thus, if an individual infected at $l+u$ has $\mathbb{E}\bigg[Z(t,l+u)(Z(t,l+u)-1)\bigg] = 0$, this individual cannot cause new infections whose infection trees have non-zero prevalence at time $l+u$. Hence, the condition \ref{eq:condition} is equivalent to the epidemic being non-expanded at time $t$, as at each time $l+u$, either no infections are possible from the initial individual, or any individuals that are infected at time $l+u$ contribute zero prevalence at time $t$ from the new infections they cause.

Hence, $Z(t,l)$ is strictly overdispersed for expanded epidemics. This means that Gaussian approximations are unlikely to be useful. 

\subsubsection*{Variance midway through an epidemic}
It is important to calculate uncertainty starting midway through an epidemic, conditional on previous events. This derivation is significantly more algebraically involved than the other work in this paper. For simplicity, we assume that $N(t,l)$ is an inhomogeneous Poisson Process, and that $L = \infty$ for each individual.

Suppose that prevalence (here equivalent to cumulative incidence) $Z(t,l) = n+1$. We create a strictly increasing sequence $l = B_0 < B_1<\cdots<B_n$ of $n+1$ infection times, which has probability density function
\begin{equation}
    \label{eq:probability_mass_function}
     \underbrace{f_{\boldsymbol{B}}(\boldsymbol{b})}_{\text{Joint pdf}} =\underbrace{\frac{1}{P(Z(t,l) = n+1)}}_{\text{normalising constant}} \underbrace{\prod_{i=1}^n\bigg(\rho(b_i) \sum_{j=0}^{i-1}\nu(b_i-b_j)\bigg)}_{\text{infection rates at each $b_i$}}\underbrace{\exp\bigg[-\sum_{i=0}^n \int_0^{t-b_i} \rho(s+l)\nu(s)ds\bigg]}_{\text{probability of no other infections}},
\end{equation}
where pdf is short for probability mass function. Then, the variance at time $t+s$ is given by
\begin{align}
    &\underbrace{\text{var}(Z(t+s,l))}_{\text{variance}} =\underbrace{\int_{b=0}^t\sum_{i=0}^nV^*(t+s,b)f_{B_i}(b)db}_{\text{variance from subsequent cases}}...\nonumber\\
    &...+ \underbrace{\int_{b=0}^t\int_{c=0}^t\sum_{i=0}^n\sum_{j=0}^nM^*(t+s,b)M^*(t+s,c)(f_{B_i,B_j}(b,c) - f_{B_i}(b)f_{B_j}(c))dbdc}_{\text{variance from unknown infection times}},
\end{align}
where $M^*(t+s,b)$ and $V^*(t+s,b)$ are the mean and variance of the size of the infection tree (i.e. prevalence or cumulative incidence) at time $t+s$, caused by an individual infected at time $b$, ignoring all individuals they infected before time $t$. These quantities are calculated from $M$ and $V$. Note also that $f_{B_i}$ and $f_{B_i,B_j}$ are the one-and-two-dimensional marginal distributions from $f_{\boldsymbol{B}}$.

\subsubsection*{Bayesian inference and for SARS epidemic in Hong Kong}

The data for the SARS epidemic in Hong Kong consist of 114 daily measurements of incidence (positive integers), and an estimate of the generation time \cite{Svensson2007-wa} obtained via the R package EpiEstim \cite{Cori2013}. We ignore the infectious period $g$ and set the infectiousness $\nu$ to the generation interval. The inferential task is then to estimate a time varying function $\rho$ from these data using Equation \ref{eq:likelihood}. As we note in Equation \ref{eq:likelihood} and in Supplementary Note 5 and 7.1-7,4, discretisation simplifies this task considerably. Our prior distributions are as follows
\begin{align*}
    \phi & \sim \text{Normal}^+(0,1)\\
    \sigma &\sim \text{Exponential(100)}\\
    \epsilon & \sim \text{Normal}(0,\sigma)\\
    \rho(t) &= \rho(t-1) + \epsilon_t
\end{align*}
where $\rho$ is modelled as a discrete random walk process. The renewal likelihood in Equation \ref{eq:likelihood} is vectorised using the approach described in \cite{Pakkanen2021-cc}. Fitting was performed in the probabilistic programming language Numpyro, using Hamiltonian Monte Carlo\cite{Hoffman2013} with 1000 warmup steps and 6000 sampling steps across two chains. The target acceptance probability was set at 0.99 with a tree depth of 15. Convergence was evaluated using the RHat statistic\cite{Gelman2003-hl}.

Forecasts were implemented through sampling using MCMC from Equation \ref{eq:likelihood}. In order to use Hamiltonian Markov Chain Monte Carlo, we relax the discrete constraint on incidence and allow it to be continuous with a diffuse prior. We ran a basic sensitivity analysis using a Random Walk Metropolis with a discrete prior to ensure this relaxation was suitable. In a forecast setting, incidence up to a time point ($T=60$) is known exactly and given as $y^{t\leq T}$. and we have access to an estimate for $\rho(t>T)$ in the future. In our case we fix $\rho(t>T)=\rho(T)$. 

Our code is available at available at \url{https://github.com/MLGlobalHealth/uncertainity_infectious_diseases.git}.

\subsubsection*{Numerically calculating the probability mass function via the probability generating function}

Following \cite{miller2018primer} and \cite{Bornemann2011-vr} (originally from \cite{Lyness1967-yk}), the probability mass function $p$ can be recovered through a pgf $F$'s derivatives at $s=0$. i.e. $\mathbb{P}(n) = \frac{1}{n!}\left(\frac{d}{ds}\right)^n F(s;t,\tau) |_{s=0}$
This is generally computationally intractable.
A well-known result from complex analysis \cite{Lyness1967-yk} holds that $ f^{(n)}(a) = \frac{n!}{2\pi i} \oint \frac{f(z)}{\left(z-a\right)^{n+1}}\, dz$ and therefore
$ \mathbb{P}(n) = \frac{1}{2\pi i}\oint\frac{F(z;t,\tau)}{z^{n+1}} dz$
This integral can be very well approximated via trapezoidal sums as
    $\mathbb{P}(n) = \frac{1}{Mr^n}\sum_{m=0}^{M-1}F(re^{2\pi i m/M};t,\tau) e^{ - 2\pi i n m/M} 
$
where $r=1$\cite{Bornemann2011-vr}. The probability mass function for any time and $n$ can be determined numerically. One needs $M\geq n$, which requires solving $n$ renewal equations for the generating function and performing a fast Fourier transform. This is computationally fast, but may become slightly burdensome for epidemics with very large numbers of infected individuals (millions). A derivation of this approximation is provided in the Supplementary Note 3.7.

\subsubsection*{Competing interests}
All authors declare no competing interests

\subsubsection*{Data availability}
 Data from Figure 3 is available via the R-Package EpiEstim\cite{Cori2013-jc}, and data from Figure 4 is available at \url{https://imperialcollegelondon.github.io/covid19local} and via official UK Government reporting (\url{https://www.ons.gov.uk/}).

\subsubsection*{Code availability}
All model code to reproduce Figures 2, 3 and 4 is available at \url{https://github.com/MLGlobalHealth/uncertainity_infectious_diseases.git}. 

\subsubsection*{Acknowledgements}

S.B.\, C.A.D\, and D.J.L.\ acknowledge support from the MRC Centre for Global Infectious Disease Analysis (MR/R015600/1), jointly funded by the UK Medical Research Council (MRC) and the UK Foreign, Commonwealth \& Development Office (FCDO), under the MRC/FCDO Concordat agreement, and also part of the EDCTP2 programme supported by the European Union. S.B. acknowledges support from the Novo Nordisk Foundation via The Novo Nordisk Young Investigator Award (NNF20OC0059309), which also supports S.M.\. S.B.\ acknowledges support from the Danish National Research Foundation via a chair position. S.B.\ and C.M.\ acknowledges support from The Eric and Wendy Schmidt Fund For Strategic Innovation via the Schmidt Polymath Award (G-22-63345). S.B.\ acknowledges support from the  National Institute for Health Research (NIHR) via the Health Protection Research Unit in Modelling and Health Economics. D.J.L.\ acknowledges funding from Vaccine Efficacy Evaluation for Priority Emerging Diseases (VEEPED) grant, (ref. NIHR:PR-OD-1017-20002) from the National Institute for Health Research. M.J.P.\ acknowledges funding from a EPSRC DTP Studentship. C.W.\ acknowledges support from the Wellcome Trust.

\subsubsection*{Author Contributions}

S.B. and M.J.P conceived and designed the study. S.B. performed analysis with assistance from M.J.P. M.J.P, D.J.L and S.B drafted the original manuscript. M.J.P drafted the supplementary information with assistance from J.P. M.J.P, D.J.L, J.P, C.W, C.M, O.R, S.M, M.P, C.A.D, and S.B revised the manuscript and contributed to its scientific interpretation. S.B, M.P and C.A.D supervised the work.

\printbibliography

\newpage
\appendix

\begin{center}
    {\LARGE \bf{Supplementary Information}}
    \newline
    \newline
    \bf{Intrinsic Randomness in Epidemic Modelling Beyond Statistical Uncertainty} 
\end{center}

\renewcommand\thefigure{\thesection.\arabic{figure}}    
\renewcommand\thetable{\thesection.\arabic{table}}    
\setcounter{figure}{0}  
\setcounter{table}{0}  
\setcounter{page}{1}

\newcommand{\beginsupplement}{%
         \setcounter{table}{0}
        \renewcommand{\thetable}{S\arabic{table}}%
        \setcounter{figure}{0}
        \renewcommand{\thefigure}{S\arabic{figure}}%
        \renewcommand \thepart{}
        \renewcommand \partname{}
     }

\beginsupplement

\part{} %

\section*{Summary}
\renewcommand{\theequation}{S.\arabic{equation}}
This supplement provides full derivations of the results from the main text. The results are, as in the main text, presented for an epidemic occurring in continuous time, although some additional results on discrete epidemics are given in the final note of this supplement. The supplement is structured as follows. 
\begin{itemize}
    \item The first note, ``Modelling'', provides a precise definition of the branching process model used throughout the paper. 
    \item The second note, ``Probability generating functions'' derives probability generating functions (pgfs) for prevalence and cumulative incidence. It also discusses their efficient solution, including some special cases in which one can speed up the solution process
    \item The third note, ``Properties of the prevalence variance'', derives the equation for the variance (via the previously derived equations for the pgf) and explores its properties, providing explanations for the various terms and proving that the prevalence of new infections is (under a mild condition on the possible spread of the epidemic) overdispersed.
    \item The fourth note, ``Likelihood functions'' contains the derivations of the pgf of the infection event times and the likelihood function presented in the main text. 
    \item The fifth note, ``Assessing future variance during an epidemic'' derives the equation for variance of future cases when the cumulative incidence is known at some point in time.
    \item Finally, the sixth note, ``Discrete epidemics'' provides a range of similar results in the discrete setting, and shows the convergence of the pgf to its continuous equivalent as the step-size tends to zero.
\end{itemize}

\section{Background literature on renewal equations}

A common approach to modelling infectious diseases is to use the renewal equation. The early theory on the properties of the renewal equation can be found here \cite{Feller1941}. Epidemiologically derived descriptions can be found here \cite{Fraser2007,Cori2013} where the renewal equation is framed in an epidemiological framework with reference to infection processes. The link between the renewal equation and the popular susceptible-infected-recovered models can be found here \cite{Champredon2018}. The basics of branching processes can be found here \cite{Harris1963-sa}. In what follows, we will arrive at a renewal equation from first principles by first starting with the probability generating function of a general branching process.

\section{Modelling}
\subsection{Branching process framework}
We present a general time-varying age-dependent branching process that is most similar to the general branching process initially proposed by Crump, Mode and Jagers \cite{Crump1968-fp,Crump1969-gc}. Following \cite{Pakkanen2021-cc}, in our process, we begin with a single individual infected at some time $l$ whose infectious period is a random variable distributed by cumulative distribution function $G(\cdot, l)$, admitting a probability density $g(\cdot,l)$. During this individual's life length, the individual gives rise to an integer-valued random number of secondary infections according to a counting processes $\{N(t,l)\}_{t\geq l}$ ($\{N(t,l)\}$ is the number of secondary infections) where $t$ is a global ``calendar'' time. The amount of time for which the individual has been infected before time $t$ is therefore $t-l$.
\\
\\
\noindent
For each infection event time - that is, for each $v$ such that
\begin{equation}
    v \in\bigg\{ u \leq t : \lim_{s \to u_{-}}(N(s,l)) \neq \lim_{s \to u_{+}}(N(s,l)) \bigg\}
\end{equation}
we then define a random variable
\begin{equation}
    Y(v,l) :=  \lim_{s \to v_{+}}(N(s,l)) - \lim_{s \to v_{-}}(N(s,l))
\end{equation}
to be the size of the infection event at time $v$; that is, this is the number of individuals that are infected (by the initial individual) at time $v$. Throughout this paper, it will be assumed that $Y = Y(v)$, so that $Y$ does not depend on the length of time for which an individual has been infected. However, this assumption could be removed from the model if desired. 
\\
\\
\noindent
Each newly infected individual then proceeds, independently, in the same way as the initial individual. The only change is that the time at which they are infected will be different (but, for example, the infection tree rooted at an individual infected at time $s > l$ is equal in distribution to the full infection tree if one started an epidemic with $l = s$). This self-similarity property underpins the derivations in the subsequent notes, as it allows an epidemic to be characterised purely by the ``first generation'' of infected individuals (and hence, the equations are derived using the ``first generation principle'').
\subsection{The counting process, \texorpdfstring{$N(t,l)$}{}}
Our framework relies on the assumption that the counting processes $N(t,l)$ has independent increments and is continuous in probability:
\begin{equation}
    \lim_{\delta \to 0}\bigg[\mathbb{P}\bigg(N(t+\delta,l) - N(t,l)\bigg)\bigg] = 0 \quad \forall t\geq l \geq 0
\end{equation}
This condition excludes any discrete formulations of the epidemic process. It will be shown later in the supplement that discrete epidemics (which are not continuous in probability), are structurally different as extra terms appear in the equations for the pgf. However, the equations in the continuous case are recovered as the step-size of the discrete process tends to zero.
\\
\\
\noindent
A further assumption on $N(t,l)$ is that it can be constructed from a L\'{e}vy Process - that is, there is some non-negative rate function $r(t,l)$ and some L\'{e}vy Process $\mathcal{N}(t)$ such that
\begin{equation}
    N(t,l) = \mathcal{N}\bigg(\int_l^tr(s,l)ds\bigg)
\end{equation}
Note that the counting processes relating to different individuals are independent, and hence will come from different independent copies of the base process $\mathcal{N}$.
\\
\\
\noindent
This assumption is important because it means that the counting process of ``infection events`` (that is, points in time such that the value of $N(t,l)$ changes) is an inhomogeneous Poisson Process, which can be shown as follows. Consider a counting process, $J_\mathcal{N}(t,l)$ that counts the increases in $\mathcal{N}$. That is,
\begin{equation}
    J_{\mathcal{N}}(t) := \bigg|\bigg\{ u \leq t : \lim_{s \to u_{-}}(\mathcal{N}(s)) \neq \lim_{s \to u_{+}}(\mathcal{N}(s)) \bigg\}\bigg|
\end{equation}
where here $|\cdot|$ denotes the number of elements in a set. Then, as $\mathcal{N}$ is a L\'{e}vy Process, $J_{\mathcal{N}}(t)$ has iid (independent and identically distributed) increments and is non-decreasing in $t$ with jumps of size 1 and thus follows a Poisson Process with some rate $\kappa$ \cite{Applebaum2009-km}. Thus, if $J(t,l)$ is the counting process of infection events in $\mathcal{N}(t,l)$, then
\begin{equation}
    J(t,l) = J_{\mathcal{N}}\bigg(\int_l^tr(s,l)ds\bigg)
\end{equation}
and hence, $J(t,l)$ is an inhomogeneous Poisson Process with rate $\kappa r(t,l)$ as required. In particular, defining 
\begin{equation}
\lambda(t,l) := \int_l^{t} r(s,l)ds,
\end{equation}
$J(t,l)$ has a generating function of
\begin{equation}
    \mathcal{J}_{(t,l)}(s) = e^{\kappa\lambda(t,l)(s-1)}
\end{equation}
\subsection{The rate function, \texorpdfstring{$r(t,l)$}{}}
Throughout the examples in this paper, the rate function $r(t,l)$ will be given as
\begin{equation}
    r(t,l) = \rho(t)\nu(t-l)
\end{equation}
Here, $\rho(t)$ is a population-level infection event rate. Note that, because the number of infections caused at each infection rate may be greater than 1 (that is one may have $J(t,l) < N(t,l)$), $\rho(t)$ cannot necessarily be interpreted in direct analogue to the reproduction number. $\nu(t-l)$ gives the infectiousness of an individual after it has been infected for time $(t-l)$. It will be assumed that $\int_0^{\infty}\nu(s)ds = 1$ so that it $\rho$ can be interpreted as the infection event rate.
\subsection{Smoothness assumptions}
Note that, throughout the derivations of this paper, the smoothness of $\rho$, $\nu$ and $g$ will not be explicitly considered when taking limits - it will be assumed that they are sufficiently smooth for ``natural'' results to hold. The authors believe that the results of this paper will hold for any piecewise continuous choices for these functions, although more detailed analysis would be needed to provide a rigorous proof of this. It is possible that they hold for much wider classes of functions, but this seems to the authors to be outside the realm of epidemiological interest, as it appears implausible that any of these functions would not be piecewise continuous in a realistic setting.
\\
\\
\noindent
Moreover, it will be assumed that unique solutions to the equations for the pgf, mean and variance exist. Again, a proof of this property is beyond the scope of this work, although the classes of equations presented in this paper are common across the literature, and it is likely that interested readers with a pure mathematical background could find applicable results to address this issue.
\subsection{Special cases for \texorpdfstring{$N(t,l)$}{}}
Throughout this paper, two special cases for $N(t,l)$ are considered - the case where $N(t,l)$ is itself an inhomogeneous Poisson Process, and the case where $N(t,l)$ is a Negative Binomial process. These were used to construct the figures in the paper and explanations as to how they can be used will be presented throughout this supplement.

\section{Probability generating functions}
\subsection{General case}
Define $F(t,l;s):=E\bigg(s^{Z(t,l)}\bigg)$ to be the generating function of $Z(t,l)$. For simplicity of notation the dependence of $F$ on $s$ will be suppressed.

To derive the generating function $F(t,l)$, we condition on  the infection period (lifetime) of the initial case, $L$.
\begin{align}
    E\bigg(s^{Z(t,l)}\bigg) &= \int_0^{\infty}E\bigg(s^{Z(t,l)}\bigg|L = u\bigg) g(u,l)du\\
    \label{eq:pgf_splitint} &= \int_{t-l}^{\infty}E\bigg(s^{Z(t,l)}\bigg|L = u\bigg) g(u,l)du + \int_0^{t-l}E\bigg(s^{Z(t,l)}\bigg|L = u\bigg) g(u,l)du
\end{align}
The counting process of the first individual, $N(t,l)$ is independent of this first individual's infection period $L$. If $L>t-l$ then this individual is still infectious and able to infect others at time $t$. Therefore, conditional on $L>t-l$, the number of people they have infected before time $t$ is independent of $L$ (as all infections from $N(s,l)_{l \leq s \leq t}$ are counted, irrespectively of the value of $L$). That is (the first term in Equation 11)
\begin{equation}
   \int_{t-l}^{\infty} E\bigg(s^{Z(t,l)}\bigg|L = u\bigg) g(u,l)du =  \int_{t-l}^{\infty} E\bigg(s^{Z(t,l)}\bigg|L\geq t-l\bigg) g(u,l)du
\end{equation}
and hence, the first integral in Supplementary Equation \ref{eq:pgf_splitint} can be simplified to give
\begin{equation}
    E\bigg(s^{Z(t,l)}\bigg)  = \bigg(1-G(t-l,l)\bigg) E\bigg(s^{Z(t,l)}\bigg|L \geq t-l\bigg) + \int_0^{t-l}E\bigg(s^{Z(t,l)}\bigg|L = u\bigg) g(u,l)du
    \label{eq:pgf_splitint_full}
\end{equation}
Let us consider the second part of Supplementary Equation \ref{eq:pgf_splitint}. Suppose first that $L = u$ for some $u < t-l$ so that the index case is no longer alive at time $t$. Thus, the number of infection events caused by the index case is given by $J(l+u,l)$. 

Define the set of times at which these infected events occurred to be $\{K_1,...,K_{J(l+u,l)}\}$ where here, importantly, the $K_i$ are labelled in a random order (so it is not necessarily the case that $K_1 < ... < K_{J(l+u,l)}$). As $J$ is an homogeneous Poisson Process and $N(t,l)$ is continuous in probability, the $K_i$ are therefore iid with pdf (probability density function)
\begin{equation}
    f_{K}(k) = \frac{r(l+k,l)}{\int_0^{u}r(l+s,l)ds}
\end{equation}
It is perhaps helpful to note that this is the step which relies on $N$ being continuous in probability. If this were not the case and $N(t,l)$ had non-zero probability of increasing at some time $s$, then the knowledge that $K_1 = s$ would give some information about $K_2$, as the fact that $K_2 \neq s$ would change its probability distribution, meaning $K_1$ and $K_2$ would not be independent. Conversely, in the continuous case, $K_1 = s$ removes an event of zero measure from the probability space of $K_2$, and hence $K_1$ and $K_2$ are still independent. 

Now, by the self-similarity property (\cite{Harris1963-sa,Kimmel1983-jo}) we have
\begin{equation}
    Z(t,l) = \sum_{i=1}^{J(l+u,l)}\sum_{j=1}^{Y(l+K_i(l+u,l))}Z_{ij}(t,l+K_i(l+u,l))
    \label{eq:self_similarity}
\end{equation}
where each $Z_{ij}$ is an independent copy of $Z$ that is equal in distribution.  $Z_{ij}$ denotes the $j$th individual corresponding to infection event time $i$. The two summations, from all previous infections, sum over all the infection events and their sizes. This summation is valid as each individual behaves independently once it has been infected.

Recall that if $X_i$ are iid random variables (with a generating function, $G_X(s)$) and if $Y$ is a non-negative integer-valued random variable (again with a generating function, $G_Y(s)$), then,
\begin{equation}
    E\bigg(s^{\sum_{i=1}^{Y}X_i}\bigg) = G_Y(G_X(s)) 
    \label{eq:pgf_lemma}
\end{equation}
By defining $\mathcal{J}_{(t,l)}$ to be the generating function of $J(t,l)$, this relationship allows us to write $\mathbb{E}(s^{Z(t,l)}|L=u)$ as
\begin{align}
    \mathbb{E}(s^{Z(t,l)}|L=u) &= \mathcal{J}_{(l+u,l)}\bigg(E\bigg[s^{\sum_{j=1}^{Y(l+K(l+u,l))}Z_j(t,l+K(l+u,l))}\bigg]\bigg)
\end{align}
where here, $K$ is equal in distribution to the $K_i$. Conditioning on the value of $K$,
\begin{equation}
    E\bigg[s^{\sum_{j=1}^{Y(l+K)}Z_j(t,l+K)}\bigg] = \int_0^{u} E\bigg[s^{\sum_{j=1}^{Y(l+k)}Z_j(t,l+k)}\bigg]\frac{r(l+k,l)}{\lambda(l+u,l)}dk
\end{equation}
Thus, defining $\mathcal{Y}_{(l+k)}$ to be the generating function of $Y(l+k)$
\begin{equation}
   E\bigg[s^{\sum_{j=1}^{Y(l+K)}Z_j(t,l+K)}\bigg] =   \int_0^{u} \mathcal{Y}_{(l+k)}(F(t,l+k))\frac{r(l+k,l)}{\lambda(l+u,l)}dk
\end{equation}
We can equivalently write this as an exponential, using the fact that $J(t,l)$ is Poisson distributed:
\begin{align}
     \mathbb{E}(s^{Z(t,l)}|L=u)&= \mathcal{J}_{(l+u,l)}\bigg(\int_0^{u}  \mathcal{Y}_{(l+k)}(F(t,l+k))\frac{r(l+k,l)}{\lambda(l+u,l)}dk\bigg)\\
     &=\exp\bigg[\kappa\lambda(l+u,l)\bigg(\int_0^{u}  \mathcal{Y}_{(l+k)}(F(t,l+k))\frac{r(l+k,l)}{\lambda(l+u,l)}dk - 1\bigg)\bigg]
\end{align}
An identical derivation can be performed on the first integral in Supplementary Equation \ref{eq:pgf_splitint} (swapping $t-l$ for $u$ and multiplying by $s$ to account for the initial case, which is counted in the prevalence at $t$ when $L > t-l$), resulting in
\begin{align}
     \mathbb{E}(s^{Z(t,l)}|L\geq t-l)&= s\mathcal{J}_{(t,l)}\bigg(\int_0^{t-l}  \mathcal{Y}_{(l+k)}(F(t,l+k))\frac{r(l+k,l)}{\lambda(t,l)}dk\bigg)\\
     &= s\exp\bigg[\kappa\lambda(t,l)\bigg(\int_0^{t-l}  \mathcal{Y}_{(l+k)}(F(t,l+k))\frac{r(l+k,l)}{\lambda(t,l)}dk -1\bigg)\bigg]
\end{align}
and therefore, this yields an overall pgf
\begin{align}
    \nonumber F(t,l) &= s\bigg(1-G(t-l,l)\bigg)\mathcal{J}_{(t,l)}\bigg(\int_0^{t-l}  \mathcal{Y}_{(l+u)}(F(t,l+u))\frac{r(l+u,l)}{\lambda(t,l)}du\bigg)... \\
    &...+\int_0^{t-l}\mathcal{J}_{(l+u,l)}\bigg(\int_0^{u}  \mathcal{Y}_{(l+k)}(F(t,l+k))\frac{r(l+k,l)}{\lambda(l+u,l)}dk\bigg)g(u,l)du
\end{align}
or, equivalently
\begin{align}
    \nonumber F(t,l) &= s\bigg(1-G(t-l,l)\bigg)\exp\bigg[\kappa\lambda(t,l)\bigg(\int_0^{t-l}  \mathcal{Y}_{(l+k)}(F(t,l+k))\frac{r(l+k,l)}{\lambda(t,l)}dk -1\bigg)\bigg]... \\
    &...+\int_0^{t-l}\exp\bigg[\kappa\lambda(l+u,l)\bigg(\int_0^{u}  \mathcal{Y}_{(l+k)}(F(t,l+k))\frac{r(l+k,l)}{\lambda(l+u,l)}dk - 1\bigg)\bigg] g(u,l) du
    \label{eq:pgf_general}
\end{align}
Note that by absorbing $\kappa$ into the rate function $r(l+k,l)$, it can be assumed that $\kappa = 1$. Intuitively this is simply scaling the probability density by the number of points.
\subsection{Solving the pgf equation}
 Practically, one will always set $l=0$ for an epidemic, and so only the values $F(t,0)$ are directly relevant. However, it is still necessary to solve for $F(t,l)$ for $0\leq l \leq t$. In the language of PDEs (partial differential equations) and, specifically, the Cauchy problem, this can be explained by the fact that the ``data curve'' is the line $t=l$ (as the values of $F(t,t)$ are known to be equal to $s$) and the ``characteristics'' of the system are the lines $t = \text{constant}$. Thus, to calculate the value of $F(t,0)$, it is necessary to follow the characteristic from $(t,t)$ to $(t,0)$ and hence calculate $F(t,l)$ for $0\leq l \leq t$.
 
Hence, following \cite{Pakkanen2021-cc}, solving Supplementary Equation \ref{eq:pgf_general} can be greatly facilitated by defining an auxiliary equation $ F_c(t) = F(c,c-t)$ and allows us to write Supplementary Equation \ref{eq:pgf_general} an equation in one variable. This is
\begin{align}
    \nonumber F_c(t) &= s\bigg(1-G(t,l)\bigg)\mathcal{J}_{(c,c-t)}\bigg(\int_0^{t}  \mathcal{Y}_{(c-t+u)}(F_c(t-u))\frac{r(c-t+u,c-t)}{\lambda(c,c-t)}du\bigg)... \\
    &...+\int_0^{t}\mathcal{J}_{(c-t+u,c-t)}\bigg(\int_0^{u}  \mathcal{Y}_{(c-t+k)}(F_c(t-k))\frac{r(c-t+k,c-t)}{\lambda(u,c-t)}dk\bigg) g(u,l) du
\end{align}
or, equivalently
\begin{align}
     \nonumber F_c(t) &= s\bigg(1-G(t,l)\bigg)\exp\bigg[\lambda(c,c-t)\kappa\bigg(\int_0^{t}  \mathcal{Y}_{(c-t+u)}(F_c(t-u))\frac{r(c-t+u,c-t)}{\lambda(c,c-t)}du-1\bigg)\bigg]... \\
    &...+\int_0^{t}\exp\bigg[\lambda(u,c-t)\kappa\bigg(\int_0^{u}  \mathcal{Y}_{(c-t+k)}(F_c(t-k))\frac{r(c-t+k,c-t)}{\lambda(u,c-t)}dk-1\bigg)\bigg]g(u,l) du
\end{align}
\subsection{Poisson case}
 If $N(t,l)$ is an inhomogeneous Poisson Process, then, as the infection event size for a Poisson Process is always 1 \cite{Applebaum2009-km}, one has $\mathcal{Y}_{(t)}(s) = s$. To aid understanding below in the Negative Binomial case, it is helpful to note that the Lévy Process, $\mathcal{N}$, can hence be characterised by
\begin{align}
\mathbb{P}( \mathcal{N}(t+dt) - \mathcal{N}(t) = 0) &= 1-\kappa dt \nonumber\\
\mathbb{P}(\mathcal{N}(t+dt) - \mathcal{N}(t) = 1) &= \kappa dt\nonumber\\
\mathbb{P}(\mathcal{N}(t+dt) - \mathcal{N}(t) > 1) &= o(dt)\nonumber
\end{align}
Setting $\kappa = 1$ as discussed above, the generating function equation becomes
\begin{align}
     F(t,l) &= s\bigg(1-G(t-l,l)\bigg)\exp\bigg[\bigg(\int_0^{t-l}  F(t,l+k)\rho(l+k)\nu(k)dk -\lambda(t,l)\bigg)\bigg]... \\
    &...+\int_0^{t-l} \exp\bigg[\bigg(\int_0^{u}  F(t,l+k)\rho(l+k)\nu(k)dk - \lambda(l+u,l)\bigg)\bigg] g(u,l) du
\end{align}
This equation can be further simplified by recalling that
\begin{align}
   \lambda(t,l) &:= \int_l^{t} r(u,l) du = \int_0^{t-l}r(u+l,l) du = \int_0^{t-l} \rho(u+l)\nu(u) du
\end{align}
therefore
\begin{align}
     F(t,l) &= s\bigg(1-G(t-l,l)\bigg)\exp\bigg[\bigg(\int_0^{t-l}  F(t,l+k)\rho(l+k)\nu(k)dk -\int_0^{t-l} \rho(l+k)\nu(k) dk\bigg)\bigg]... \nonumber\\
    &...+\int_0^{t-l} \exp\bigg[\bigg(\int_0^{u}  F(t,l+k)\rho(l+k)\nu(k)dk - \int_0^{u} \rho(l+k)\nu(k) du\bigg)\bigg] g(u,l) du \nonumber\\
    \nonumber\\
     &= s\bigg(1-G(t-l,l)\bigg)\exp\bigg[\bigg(\int_0^{t-l}  \rho(l+k)\nu(k) \left(F(t,l+k)-1\right)dk \bigg)\bigg]... \nonumber\\
    &...+\int_0^{t-l} \exp\bigg[\bigg(\int_0^{u}  \rho(l+k)\nu(k)dk \left(F(t,l+k)-1\right) \bigg)\bigg] g(u,l) du 
\end{align}

For computational ease the auxiliary function equation is then
\begin{align}
        F_c(t) &=s \bigg(1-G(t,l)\bigg)  \text{exp}\bigg[\bigg(\int_0^{t} \left(F_c(t-u)-1\right)\rho(c-t+u)\nu(u)du \bigg)\bigg]...    \nonumber 
        \\
    &...+\int_0^{t}\text{exp}\bigg[\bigg(\int_0^u \left(F_c(t-k)-1\right)\rho(c-t+k)\nu(k)dk \bigg)\bigg] g(u,l)du
\end{align}
\subsection{Inhomogeneous Negative Binomial case}
Our derivation follows from the well-known relationship that the Negative Binomial distribution arises from a compound Poisson distribution. For $p\in(0,1)$ and $\phi\in\mathbb{R}^+$, if 
\begin{equation}
    X=\sum_{i=1}^N Y_i
\end{equation}
where 
\begin{equation}
    N\sim\text{Poisson}(-\phi\ln(p))
\end{equation}
and each $Y_i$ is independent of $N$, iid, and follows a logarithmic series distribution
\begin{equation}
    Y_i\sim\text{Logarithmic}(1-p)
\end{equation}
then the random variable $X$ is Negative Binomial distributed. This can easily be proven using pgfs. Therefore we have  $\kappa=-\ln(p)\phi$ and can calculate the pgf for $Y$ as $\mathcal{Y}(s) =  \frac{\ln(1-(1-p)s)}{\ln(p)}$. These can then be substituted into our general Supplementary Equation \ref{eq:pgf_general}. 

For clarity we re-derive this relationship explicitly. We have
\begin{equation}
    \mathcal{N}(t) \sim \text{NB}(\phi t, p)
\end{equation}
As $M(t)$ has iid increments,
\begin{equation}
    \mathbb{P}\bigg(\mathcal{N}(t+dt) - \mathcal{N}(t) = k\bigg) =  \mathbb{P}\bigg(\mathcal{N}(dt)= k\bigg) = \frac{(k+\phi dt-1)(k+\phi dt-2)...\phi dt}{k!}(1-p)^kp^{\phi dt}
\end{equation}
Thus, to leading order, for $k > 0$, one has
\begin{equation}
    \mathbb{P}\bigg(\mathcal{N}(t+dt) - \mathcal{N}(t) = k\bigg) = \frac{(1-p)^k\phi dt}{k} + o(dt)
\end{equation}
while if $k=0$,
\begin{equation}
     \mathbb{P}\bigg(\mathcal{N}(t+dt) - \mathcal{N}(t) = 0\bigg) = p^{\phi dt} = 1 + \ln(p)\phi dt + o(dt)
\end{equation}
(noting that $\ln(p) < 0$). This means that the infection event process $J_{\mathcal{N}}$ satisfies
\begin{equation}
    \mathbb{P}\bigg(J_{\mathcal{N}}(t+dt) - J_{\mathcal{N}}(t) = 0\bigg) =  1 + \ln(p)\phi dt + o(dt)
\end{equation}
and
\begin{align}
     \mathbb{P}\bigg(J_{\mathcal{N}}(t+dt) - J_{\mathcal{N}}(t) = 1\bigg) &=  \sum_{k=1}^{\infty}\frac{(1-p)^k\phi dt}{k} + o(dt)\\
     &= - \ln(p)\phi dt + o(dt)
\end{align}
and hence, $J_{\mathcal{N}}$ is a Poisson Process of rate $-\ln(p)\phi$ \cite{Barndorff-Nielsen1969-vx} . Thus, one has 
\begin{equation}
    \kappa = -\ln(p)\phi
\end{equation}
as expected. Moreover, the pmf (probability mass function) of a infection event size, $Y$ is given by
\begin{equation}
\label{eq:logpdf}
    \mathbb{P}(Y = k) = \frac{(1-p)^k }{- k \ln(p)}
\end{equation}
One can hence find the generating function as
\begin{equation}
    \mathcal{Y}(s) = \sum_{k=1}^{\infty} \frac{((1-p)s)^k \phi}{-k \ln(p)}
\end{equation}
Noting that
\begin{equation}
     \sum_{k=1}^{\infty} \frac{(1-p)^k}{-k \ln(p)} = 1
\end{equation}
one has
\begin{equation}
     \mathcal{Y}(s) =  \frac{\ln(1-(1-p)s)}{\ln(p)}\sum_{k=1}^{\infty} \frac{(1 - (1 - (1-p)s))^k }{-k \ln(1-(1-p)s)} =  \frac{\ln(1-(1-p)s)}{\ln(p)}
\end{equation}
These results can be substituted into the general formula to give
\begin{align}
     F(t,l) = & s\bigg(1-G(t-l,l)\bigg)\exp\bigg[ -\phi \bigg(\int_0^{t-l}\ln(1-(1-p)F(t,l+u))  \rho(u+l)\nu(u)du  + \ln(p)\lambda(t,l)\bigg)\bigg]...\nonumber\\
     &...+   \int_0^{t-l}\exp\bigg[ -\phi\bigg(\int_0^{u}\ln(1-(1-p)F(t,l+k))  \rho(k+l)\nu(k)dk  + \ln(p)\lambda(u,l)\bigg)\bigg]g(u,l)du
\end{align}
As in the Poisson case, this equation can be simplified by factoring $\lambda$ \begin{align}
     F(t,l) = & s\bigg(1-G(t-l,l)\bigg)\exp\bigg[ -\phi \bigg(\int_0^{t-l}\left(\ln(1-(1-p)F(t,l+u)) - \ln(p) \right)  \rho(u+l)\nu(u)du \bigg)\bigg]...\nonumber\\
     &...+   \int_0^{t-l}\exp\bigg[ -\phi\bigg(\int_0^{u} \left(\ln(1-(1-p)F(t,l+k)) -\ln(p)\right) \rho(k+l)\nu(k)dk  \bigg)\bigg]g(u,l)du
\end{align}
The easier-to-solve auxiliary function is given by
\begin{align}
      F_c(t) = &s\bigg(1-G(t-l,l)\bigg)\exp\bigg[ -\phi \bigg(\int_0^{t}\left(\ln(1-(1-p)F_c(t-u)) -\ln(p)\right) \rho(c-t+u)\nu(u)du  \bigg)\bigg]...\nonumber\\
      &...+    \int_0^{t}\exp\bigg[ -\phi  \bigg(\int_0^{u}\left(\ln(1-(1-p)F_c(t-k)) -\ln(p)\right)  \rho(c-t+k)\nu(k)dk  \bigg)\bigg]g(u,l)du
      \label{eq:pgf_negativebin_auxiliary}
\end{align}
If $p = \frac{\phi}{1 + \phi}$, then the Poisson case (with $\kappa = 1$) is recovered in the $\phi \to \infty$ limit.

Note that $\mathbb{E}[N(t,l)] = \frac{\phi \lambda(t,l) (1-p)}{p}$ while in our case, we impose that $\mathbb{E}[N(t,l)] = \lambda(t,l)$. Solving for $p$ we can see $p = \frac{\phi}{1+\phi}$ and this relation can be substituted into Supplementary Equation \ref{eq:pgf_negativebin_auxiliary}. Note that this agrees with the definition of $p$ in the Poisson limit.
\subsection{Cumulative incidence}

Similar to prevalence, cumulative incidence can be calculated by counting all previous infections as well as current ones. Following an identical derivation to prevalence the pgf for cumulative incidence simply requires multiplying the second integral by $s$ as the initial infection is counted in the cumulative incidence regardless of the value of $L$.
\begin{align}
    \nonumber F(t,l) &= s\bigg(1-G(t-l,l)\bigg)\mathcal{J}_{(t,l)}\bigg(\int_0^{t-l}  \mathcal{Y}_{(l+u)}(F(t,l+u))\frac{r(l+u,l)}{\lambda(t,l)}du\bigg)... \\
    &...+s\int_0^{t-l}\mathcal{J}_{(l+u,l)}\bigg(\int_0^{u}  \mathcal{Y}_{(l+k)}(F(t,l+k))\frac{r(l+k,l)}{\lambda(t,l)}dk\bigg)g(u,l)du
\end{align}

\subsection{A simplified pgf ignoring \texorpdfstring{$g$}{}}
By assuming $g(u,l)=0 ~~\forall~~ u$ and therefore $G(u,l)=0~~ \forall ~~u$, the pgf for prevalence (or, in this case, equivalently, cumulative incidence) simplifies to
\begin{align}
    \nonumber F(t,l) &= s\mathcal{J}_{(t,l)}\bigg(\int_0^{t-l}  \mathcal{Y}_{(l+u,l)}(F(t,l+u))\frac{r(l+u,l)}{\lambda(t,l)}du\bigg)
    \label{eq:pdf_incidence_approx}
\end{align}
Additional computational savings can be gained in our case $r(t,l)=\rho(t)\nu(t-l)$ if the infectiousness $\nu$ decays to zero quickly. This means that the auxiliary equation used for computation can be truncated to some time $\text{min}(t,T)$. For example, in the Poisson case this becomes,
\begin{align}
        F_c(t) &= \exp\bigg[\bigg(\int_0^{\text{min}(t,T)}  \left(F_c(t-u)-1\right)\rho(c-t+u)\nu(u)du \bigg)\bigg]s 
\end{align}
and in the Negative Binomial case this becomes,
\begin{align}
      F_c(t) = & \exp\bigg[ -\phi \bigg(\int_0^{\text{min}(t,T)} \left(\ln(1-(1-p)F_c(t-u)) -\ln(p)\right) \rho(c-t+u)\nu(u)du  \bigg)\bigg]s
\end{align}
These computational savings allow computation of the pgf for millions of iterations in minutes. 
\subsection{Calculating the probability mass function via the pgf}
Following \cite{miller2018primer} and \cite{Bornemann2011-vr} (originally from \cite{Lyness1967-yk}), by the properties of pgfs, the probability mass function $p$ can be recovered through a pgf $F$'s derivatives at $s=0$
\begin{align*}
    \mathbb{P}(n) = \frac{1}{n!}\left(\frac{d}{ds}\right)^n F(s;t,\tau) |_{s=0}
\end{align*}
This is generally computationally intractable.
A well-known result from complex analysis \cite{Lyness1967-yk} holds that
\begin{align}
    f^{(n)}(a) = \frac{n!}{2\pi i} \oint \frac{f(z)}{\left(z-a\right)^{n+1}}\, dz.
\end{align}
Therefore
\begin{align}
    \mathbb{P}(n) = \frac{1}{2\pi i}\oint\frac{F(z;t,\tau)}{z^{n+1}} dz
\end{align}
This integral can be done on a closed circle around the origin such that $z=re^{i\theta}$ and $dz=izd\theta$ - i.e.
\begin{align}
    \mathbb{P}(n) = \frac{1}{2\pi}\int_0^{2\pi}\frac{F(re^{i\theta};t,\tau)}{(re^{i\theta})^n} d\theta
\end{align}
Finally through substitution $\theta = 2\pi u$ such that $d\theta=2\pi du$, where $u\in{[0,1]}$ we find
\begin{align}
    \mathbb{P}(n) = \int_0^{1}\frac{F(re^{2\pi i u};t,\tau)}{r^n e^{2\pi i u n}} du
\end{align}
Since trapezoidal sums are known to converge geometrically for periodic analytic
functions (Davis 1959) a simple approximation becomes
\begin{align}
    \mathbb{P}(n) = \frac{1}{Mr^n}\sum_{m=0}^{M-1}F(re^{2\pi i m/M};t,\tau) e^{ - 2\pi i n m/M} 
\end{align}
Bornemann\cite{Bornemann2011-vr} suggest using $r=1$.

The probability mass function for any time and $n$ can be determined numerically. One needs $M\geq n$, which requires solving $n$ renewal equations for the generating function and performing a fast Fourier transform. This is generally computationally fast, but may become slightly burdensome for epidemics with very large numbers of infected individuals. 

\section{Properties of the prevalence variance}
\subsection{Derivation of equation for mean prevalence}
Before deriving the equation for the prevalence variance, it is important to derive the equation governing the mean prevalence. This has been previously derived in \cite{Pakkanen2021-cc}, although here, we re-derive it from our new pgfs. First note that
\begin{align}
    &\frac{\partial}{\partial s}\bigg(\mathcal{J}_{(t,l)}\bigg(\int_0^{t-l}  \mathcal{Y}_{(l+u,l)}(F(t,l+u))\frac{r(l+u,l)}{\lambda(t,l)}du\bigg)\bigg)...\nonumber\\
    &= \bigg[\int_0^{t-l} F_s(t,l+u)\frac{r(l+u,l)}{\lambda(t,l)} \mathcal{Y}_{(l+u,l)}'(F(t,l+u))du\bigg]\bigg[\mathcal{J}'_{(t,l)}\bigg(\int_0^{t-l}  \mathcal{Y}_{(l+u,l)}(F(t,l+u))\frac{r(l+u,l)}{\lambda(t,l)}du\bigg)\bigg)\bigg]
\end{align}
Now, setting $s = 1$ so that $F(\cdot,\cdot) = 1$ and $F_s(\cdot,\cdot) = M(\cdot,\cdot)$, one has
\begin{equation}
    \bigg[\int_0^{t-l} M(t,l+u)\frac{r(l+u,l)}{\lambda(t,l)} \mathcal{Y}_{(l+u,l)}'(1)du\bigg]\bigg[\mathcal{J}'_{(t,l)}\bigg(\int_0^{t-l}  \mathcal{Y}_{(l+u,l)}(1)\frac{r(l+u,l)}{\lambda(t,l)}du\bigg)\bigg)\bigg]
\end{equation}
Now, define $B(t) = \mathbb{E}(Y(t))$ so that $\mathcal{Y}_{(l+u)}'(1) = B(l+u)$. Moreover, $\mathcal{Y}_{(l+u)}(1) = 1$ so the equation becomes
\begin{equation}
     \bigg[\int_0^{t-l} M(t,l+u)\frac{r(l+u,l)}{\lambda(t,l)} B(l+u)du\bigg]\bigg[\mathcal{J}'_{(t,l)}\bigg(\int_0^{t-l} \frac{r(l+u,l)}{\lambda(t,l)}du\bigg)\bigg)\bigg]
\end{equation}
Now, necessarily
\begin{equation}
    \int_0^{t-l} \frac{r(l+u,l)}{\lambda(t,l)}du = 1 \Rightarrow \mathcal{J}'_{(t,l)}\bigg(\int_0^{t-l} \frac{r(l+u,l)}{\lambda(t,l)}du\bigg) = \mathbb{E}(J(t,l)) = \kappa \lambda(t,l)
\end{equation}
and so, this results in
\begin{equation}
    \int_0^{t-l} M(t,l+u)\frac{r(l+u,l)}{\lambda(t,l)} B(l+u)\kappa \lambda(t,l)du
\end{equation}
Moreover, evaluating 
\begin{equation}
    \mathcal{J}_{(t,l)}\bigg(\int_0^{t-l}  \mathcal{Y}_{(l+u)}(F(t,l+u))\frac{r(l+u,l)}{\lambda(t,l)}du\bigg)
\end{equation}
at $s=1$ gives
\begin{align}
    \mathcal{J}_{(t,l)}\bigg(\int_0^{t-l}  \mathcal{Y}_{(l+u)}(1)\frac{r(l+u,l)}{\lambda(t,l)}du\bigg) &=   \mathcal{J}_{(t,l)}\bigg(\int_0^{t-l} 1\times \frac{r(l+u,l)}{\lambda(t,l)}du\bigg)\\
    &= \mathcal{J}_{(t,l)}(1)\\
    &=1
\end{align}
Thus, the derivative of the full generating function equation gives
\begin{align}
\label{eq:meanold}
    M(t,l) &= (1-G(t-l,l))\bigg[1 + \int_0^{t-l} M(t,l+u)\frac{r(l+u,l)}{\lambda(t,l)} B(l+u)\kappa \lambda(t,l)du\bigg]...\\
    &...+ \int_0^{t-l}\int_0^uM(t,l+k)\frac{r(l+k,l)}{\lambda(l+u,l)}B(l+k)\kappa \lambda(l+u,l)g(u,l) dk du
\end{align}

This can be simplified significantly. Note that,
\begin{align}
    &\nonumber \int_0^{t-l}\int_0^uM(t,l+k)\frac{r(l+k,l)}{\lambda(l+u,l)}B(l+k)\kappa \lambda(l+u,l)g(u,l) dk du =\\ &\int_0^{t-l}\int_0^uM(t,l+k)r(l+k,k)B(l+k)\kappa g(u,l) dk du
\end{align}
Moreover, one can change the order of integration to get
\begin{equation}
    \int_0^{t-l}\int_k^{t-l}M(t,l+k)r(l+k,k)B(l+k)\kappa g(u,l) du dk = \int_0^{t-l}M(t,l+k)r(l+k,k)B(l+k)\kappa (G(t-l,l) - G(k,l))
\end{equation}
and hence, one can write the equation for $M(t,l)$ as
\begin{equation}
\label{eq:Meaneq}
    M(t,l) = (1-G(t-l,l)) + \int_0^{t-l} M(t,l+u)r(l+u,l) \left(B(l+u)\kappa\right) (1-G(u,l))du
\end{equation}
Note that, for the Poisson special case, $B(l+u,l) = 1$ and for the Negative Binomial special case, $B(l+u,l) = \frac{p-1}{p\ln(p)} = -\frac{1}{\ln(p)\phi}$. In both cases, it may improve the epidemiological interpretation of $\rho$ to absorb the $B(l+u,l)\kappa$ term into $\rho$ (so that $\rho$ becomes a measure of the rate of new infections). This gives the simpler equation
\begin{equation}
    M(t,l) = (1-G(t-l,l)) + \int_0^{t-l} M(t,l+u)\rho(l+u)\nu(u) (1-G(u,l))du
\end{equation}
which agrees with \cite{Pakkanen2021-cc}. 
\subsection{Derivation of equation for prevalence variance}
The equation for variance can now be found by taking the second derivative of the pgf. Define $W(t,l) := \mathbb{E}(Z(t,l)(Z(t,l) - 1))$. Note that this then gives the variance, $V(t,l)$ as $V(t,l) = W(t,l)+M(t,l) - M(t,l)^2$.
\\
\\
\noindent
Consider first the term
\begin{equation}
    s\bigg(1-G(t-l,l)\bigg)\mathcal{J}_{(t,l)}\bigg(\int_0^{t-l}  \mathcal{Y}_{(l+u)}(F(t,l+u))\frac{r(l+u,l)}{\lambda(t,l)}du\bigg)
\end{equation}
The first derivative of this term is equal to
\begin{align}
    &\bar{G}(t-l,l)\mathcal{J}_{(t,l)}\bigg(\int_0^{t-l}  \mathcal{Y}_{(l+u)}(F(t,l+u))\frac{r(l+u,l)}{\lambda(t,l)}du\bigg) + ...\nonumber \\
    &s\bar{G}(t-l,l)\bigg[\int_0^{t-l}F_s(t,l+u)\mathcal{Y}'_{(l+u)}(F(t,l+u))\frac{r(l+u,l)}{\lambda(t,l)}du\bigg]\mathcal{J}'_{(t,l)}\bigg(\int_0^{t-l}  \mathcal{Y}_{(l+u)}(F(t,l+u))\frac{r(l+u,l)}{\lambda(t,l)}du\bigg)
\end{align}
Then, the second derivative is equal to
\begin{align}
    &2\bar{G}(t-l,l)\bigg[\int_0^{t-l}F_s(t,l+u)\mathcal{Y}'_{(l+u)}(F(t,l+u))\frac{r(l+u,l)}{\lambda(t,l)}du\bigg]\mathcal{J}'_{(t,l)}\bigg(\int_0^{t-l}  \mathcal{Y}_{(l+u)}(F(t,l+u))\frac{r(l+u,l)}{\lambda(t,l)}du\bigg) +\nonumber\\
    &+s\bar{G}(t-l,l)\bigg[\int_0^{t-l}F_s(t,l+u)\mathcal{Y}'_{(l+u)}(F(t,l+u))\frac{r(l+u,l)}{\lambda(t,l)}du\bigg]^2\mathcal{J}''_{(t,l)}\bigg(\int_0^{t-l}  \mathcal{Y}_{(l+u)}(F(t,l+u))\frac{r(l+u,l)}{\lambda(t,l)}du\bigg)\nonumber\\
    &+s\bar{G}(t-l,l)\bigg[\int_0^{t-l}F_{ss}(t,l+u)\mathcal{Y}'_{(l+u)}(F(t,l+u))\frac{r(l+u,l)}{\lambda(t,l)}du\bigg]\mathcal{J}'_{(t,l)}\bigg(\int_0^{t-l}  \mathcal{Y}_{(l+u)}(F(t,l+u))\frac{r(l+u,l)}{\lambda(t,l)}du\bigg)\nonumber\\
    &+s\bar{G}(t-l,l)\bigg[\int_0^{t-l}F^2_s(t,l+u)\mathcal{Y}''_{(l+u)}(F(t,l+u))\frac{r(l+u,l)}{\lambda(t,l)}du\bigg]\mathcal{J}'_{(t,l)}\bigg(\int_0^{t-l}  \mathcal{Y}_{(l+u)}(F(t,l+u))\frac{r(l+u,l)}{\lambda(t,l)}du\bigg)
\end{align}
Now, one can evaluate this as $s=1$. Note that
\begin{align}
    \int_0^{t-l}  \mathcal{Y}_{(l+u)}(F(t,l+u))\frac{r(l+u,l)}{\lambda(t,l)}du &= \int_0^{t-l}  \mathcal{Y}_{(l+u)}(1)\frac{r(l+u,l)}{\lambda(t,l)}du\nonumber\\
    &= \int_0^{t-l} 1 \times \frac{r(l+u,l)}{\lambda(t,l)}du\nonumber\\
    &= 1
\end{align}
Moreover, define $B^W(t) := \mathbb{E}(Y(t)(Y(t) - 1))$ and $C^W(t,l) := \mathbb{E}(J(t,l)(J(t,l)-1))$. Note also $\mathbb{E}(J(t,l))=\lambda(t,l)$. Thus, the second derivative evaluated at $s=1$ is
\begin{align}
\label{eq:cumincdifference} &2\bar{G}(t-l,l)\bigg[\int_0^{t-l}M(t,l+u)B(l+u)\frac{r(l+u,l)}{\lambda(t,l)}du\bigg]\kappa \lambda(t,l)\nonumber\\
    &+\bar{G}(t-l,l)\bigg[\int_0^{t-l}M(t,l+u)B(l+u)\frac{r(l+u,l)}{\lambda(t,l)}du\bigg]^2C^W(t,l\nonumber)\\
    &+ \bar{G}(t-l,l)\bigg[\int_0^{t-l}W(t,l+u)B(l+u)\frac{r(l+u,l)}{\lambda(t,l)}du\bigg]\kappa \lambda(t,l)\nonumber\\
    &+  \bar{G}(t-l,l)\bigg[\int_0^{t-l}M(t,l+u)^2B^W(l+u)\frac{r(l+u,l)}{\lambda(t,l)}du\bigg]\kappa \lambda(t,l)
\end{align}
Noting that $J(t,l)$ is Poisson, one has
\begin{equation}
    C^W(t,l) + \mathbb{E}(J(t,l)) - \mathbb{E}(J(t,l)^2) = \text{var}(J(t,l)) = \mathbb{E}(J(t,l))
\end{equation}
and hence
\begin{equation}
    C^W(t,l) = \mathbb{E}(J(t,l))^2
\end{equation}
Define
\begin{align}
    \chi(t,l,k) :=\kappa \bigg[W(t,l+k)B(l+k)r(l+k,l)+ M(t,l+k)^2B^W(l+k)r(l+k,l)\bigg]
\end{align}
Then, the same process can be carried out for the second part of the equation to give
\begin{align}
    W(t,l) = &2\bar{G}(t-l,l)\bigg[\int_0^{t-l}M(t,l+u)B(l+u)r(l+u,l)du\bigg] + \bar{G}(t-l,l)\int_0^{t-l}\chi(t,l,k)dk...\nonumber\\
    ...&+\bar{G}(t-l,l)\bigg[\int_0^{t-l}M(t,l+u)B(l+u)\kappa r(l+u,l)du\bigg]^2...\nonumber\\
    ...&+ \int_0^{t-l}\bigg[\int_0^{u}M(t,l+u)B(l+u)\kappa r(l+u,l)du\bigg]^2g(u,l)du...\nonumber\\
    ...& +  \int_0^{t-l}\int_0^u\chi(t,l,k)dkg(u,l)du
\end{align}
For ease of notation, define
\begin{equation}
    S(t,l,u) := \bigg[\int_0^{u}M(t,l+k)B(l+k)\kappa r(l+k,l)dk\bigg]^2
\end{equation}
so that
\begin{align}
    W(t,l) = &2\bar{G}(t-l,l)\bigg[\int_0^{t-l}M(t,l+u)B(l+u)r(l+u,l)du\bigg] + \bar{G}(t-l,l)\int_0^{t-l}\chi(t,l,k)dk...\nonumber\\
    ...&+\bar{G}(t-l,l)S(t,l,t-l) + \int_0^{t-l}S(t,l,u)g(u,l)du+  \int_0^{t-l}\int_0^u\chi(t,l,k)dkg(u,l)du
\end{align}
Now, changing the order of integration in the final term (as was done in the derivation of the mean prevalence), this can be rewritten as 
\begin{align}
    W(t,l) &= 2\bar{G}(t-l,l)\bigg[\int_0^{t-l}\kappa M(t,l+u)B(l+u)r(l+u,l)du\bigg] +\bar{G}(t-l,l)S(t,l,t-l)...\nonumber\\ 
    &...+ \int_0^{t-l}S(t,l,u)g(u,l)du+ \int_0^{t-l}\chi(t,l,k)\bar{G}(k,l)dk
\end{align}
From this, we can create an equation for $\mathbb{E}(Z(t,l)^2) := X(t,l) = W(t,l) + M(t,l)$ by defining
\begin{equation}
    \chi^X(t,l,k) = \kappa \bigg[X(t,l+k)B(l+k,l)r(l+k,l)+ M(t,l+k)^2B^W(l+k,l)r(l+k,l)\bigg]
\end{equation}
and then simply adding the equation for $M$ to give
\begin{align}
     X(t,l) &=\bar{G}(t-l,l)+ 2\bar{G}(t-l,l)\bigg[\int_0^{t-l}\kappa M(t,l+u)B(l+u)r(l+u,l)du\bigg] +\bar{G}(t-l,l)S(t,l,t-l)...\nonumber\\ 
    \label{eq:meanadd} &...+ \int_0^{t-l}S(t,l,u)g(u,l)du+ \int_0^{t-l}\chi^X(t,l,k)\bar{G}(k,l)dk
\end{align}
Finally, to form the equation for the variance $V(t,l) = X(t,l) - M(t,l)^2$, note that
\begin{align}
    \chi^X(t,l,k) &= \kappa \bigg[X(t,l+k)B(l+k)r(l+k,l)+ M(t,l+k)^2(\mathbb{E}(Y(l+k)^2) -  B(l+k))r(l+k,l)\bigg]\\
    &= \kappa\bigg[V(t,l+k)B(l+k)r(l+k,l)+ M(t,l+k)^2\mathbb{E}(Y(l+k)^2)r(l+k,l)\bigg]\\
    &:= \chi^V(t,l,k)
\end{align}
and hence, subtracting $M(t,l)^2$ from both sides of the equation for $X(t,l)$ gives
\begin{align}
\label{eq:Vareq}
    V(t,l) &=\bar{G}(t-l,l)+ 2\bar{G}(t-l,l)\bigg[\int_0^{t-l}\kappa M(t,l+u)B(l+u)r(l+u,l)du\bigg] +\bar{G}(t-l,l)S(t,l,t-l)...\nonumber\\ 
    &...+ \int_0^{t-l}S(t,l,u)g(u,l)du+ \int_0^{t-l}\chi^V(t,l,k)\bar{G}(k,l)dk - M(t,l)^2
\end{align}
\subsection{An explanation of the variance equation}
There are two main sources of uncertainty in the infection process - the infectious period of an individual, and the number and timing of infections that occur during this infectious period. One can show that the variance splits into three terms - one for each of these two sources of uncertainty from the initial individual, and one which propagates the uncertainty through the descendants of the initial individual.
\\
\\
\noindent
Each term will be derived by assuming that all other parts of the model are deterministic. To begin, suppose that the infectious period of the initial individual is random but all other parts of the model are deterministic, so that, given that the initial individual is infectious at time $l+u$, it will infect $B(l+u)r(l+u,l)dt$ people in the interval $[u,u+dt]$ (note that this is an abstraction to illustrate the source of this variance, as it is impossible for non-integer numbers of infections to occur). Moreover, it is assumed that each of these individuals have given rise to exactly $M(t,l+u)$ infections at time $t$. Then, note that
\begin{align}
    \text{var}(Z(t,l)) &= \mathbb{E}(Z(t,l)^2) - \mathbb{E}(Z(t,l))^2\\
    &= \int_0^{\infty}\mathbb{E}(Z(t,l)^2 | L = u)g(u,l)du -M(t,l)^2\\
    &= \int_0^{t-l}\bigg[\int_0^{u}M(t,l+k)B(l+k)r(l+k,l)dk\bigg]^2g(u,l)du ...\\
    \nonumber&...+\bar{G}(t-l,l)\bigg(1 + \int_0^{t-l}M(t,l+k)B(l+k)r(l+k,l)dk\bigg)^2 - M(t,l)^2\\
    &=  \bar{G}(t-l,l) + 2\bar{G}(t-l,l)\int_0^{t-l}M(t,l+k)B(l+k)r(l+k,l)dk +...\\
    \nonumber&...+\bar{G}(t-l,l)S(t,l,t-l) + \int_0^{t-l}S(t,l,u)g(u,l)du - M(t,l)^2
\end{align}
which recovers all the terms of the variance equation except for $\int_0^{t-l}\chi^V(t,l,k)\bar{G}(k,l)dk$. 
\\
\\
\noindent
Now, suppose that the infectious period of the initial individual is deterministic in the sense that they infect others at a rate of $r(l+k,l)\bar{G}(k,l)$, i.e. the expected rate at time $l+k$. Thus, the number of infection events in the interval $[l+(k-1)dt,l+kdt]$ is (to leading order in $dt$) a Poisson variable, $A_k$, with mean $r(l+k,l)\bar{G}(k,l)dt$ and hence the number of infections is that Poisson variable multiplied by $Y(l+k,l)$. Finally, note that, as before, any individuals born at time $l+k$ will be assumed to deterministically cause $M(t,l+k)$ active infections at time $t$. Thus,
\begin{align}
    \text{var}(Z(t,l)) &= \int_{k=0}^{k=t-l}\text{var}(M(t,l+k)Y(l+k)A_k)\\
    &= \int_0^{t-l}\mathbb{E}((M(t,l+k)Y(l+k)A_k)^2) - \int_{k=0}^{k=t-l}\mathbb{E}((M(t,l+k)Y(l+k)A_k))^2\\
    &=\int_{k=0}^{k=t-l}M(t,l+k)^2\mathbb{E}(Y(l+k)^2)\mathbb{E}(A_k^2) -\int_{k=0}^{k=t-l}B(l+k)^2r(l+k,l)^2\bar{G}(k,l)^2dt^2M(t,l+k)^2
\end{align}
Ignoring the $dt^2$ term as it has zero measure, and noting that $Y$ and $A_k$ are independent
\begin{equation}
     \text{var}(Z(t,l)) = \int_0^{t-l}M(t,l+k)^2\mathbb{E}(Y(l+k,l)^2)r(l+k,l)\bar{G}(k,l)dt
\end{equation}
which is again a term from the variance equation.
\\
\\
\noindent
The final term, $\int_0^{t-l}V(t,l+k)B(l+k)\bar{G}(k,l)r(l+k,l)dk$ denotes the propagation of uncertainty through future generations. Indeed, if the infection process of the initial individual (and its infectious period) are assumed to be fully deterministic, then one simply has
\begin{equation}
    \text{var}(Z(t,l)) =\int_0^{t-l}\text{var}(Z(t,l+k))\mathbb{E}(\text{number of individuals born at $l+k$})
\end{equation}
which can easily be seen to give the correct term. 
\subsection{Overdispersion}
For the purposes of this note, it is helpful to create the following definition

\textbf{Expanded}: An epidemic is called ``expanded'' at time $t$, if there is a non-zero probability that the prevalence, not counting the initial individual or its secondary infections, is non-zero.

In this note, it will be shown that, if $\tilde{Z}(t,l)$ is the prevalence of \emph{new} infections (that is, the prevalence without counting the initial case) then if the epidemic is expanded at time $t$, $\tilde{Z}(t,l)$ is strictly overdispersed. That is
\begin{equation}
    \text{var}(\tilde{Z}(t,l)) > \mathbb{E}(\tilde{Z}(t,l)) \quad \text{or} \quad \mathbb{E}(\tilde{Z}(t,l+k))\rho(l+k,l)\nu(k)\bar{G}(k,l) = 0 \quad \forall k \in (0,t-l)
\end{equation}
The second condition ensures that, at each $k$, either the likelihood of a new infection being caused at time $l+k$, or the probability of an individual who was infected at time $l+k$ causing subsequent infections whose infection tree has non-zero prevalence at time $t$, is zero. Hence, it is equivalent to the epidemic not being expanded at time $t$.

It is crucial to use $\tilde{Z}(t,l)$ rather than $Z(t,l)$, as otherwise the deterministic initial case means that, for early times, the prevalence is underdispersed (as, for example $\mathbb{E}(Z(l,l)) = 1$ and $\text{var}(Z(l,l)) = 0$). Moreover, the condition on the tertiary infections is necessary as, otherwise, if $N(t,l)$ is Poissonian, then $\tilde{Z}(t,l)$ is also Poissonian (and therefore not strictly overdispersed).
\\
\\
\noindent
It is helpful to derive equations for the quantities for the mean $\tilde{M}(t,l)$ and the variance $\tilde{V}(t,l)$ of the new infection prevalence. This can be done by following the methods of the previous note. The derivations are mostly identical, and so will not be covered in detail. However, the key point is to note that the equation for the pgf, $\tilde{F}$, becomes
\begin{align}
    \nonumber \tilde{F}(t,l) &= \bigg(1-G(t-l,l)\bigg)\mathcal{J}_{(t,l)}\bigg(\int_0^{t-l}  \mathcal{Y}_{(l+u)}(\tilde{F}(t,l+u))\frac{r(l+u,l)}{\lambda(t,l)}du\bigg)... \\
    &...+\int_0^{t-l}\mathcal{J}_{(l+u,l)}\bigg(\int_0^{u}  \mathcal{Y}_{(l+k)}(\tilde{F}(t,l+k))\frac{r(l+k,l)}{\lambda(l+u,l)}dk\bigg)g(u,l)du
\end{align}
as the factor of $s$ in the first term is discarded. This equation can then be differentiated as before to show that
\begin{align}
   \nonumber \tilde{M}(t,l) &= (1-G(t-l,l))\bigg[\int_0^{t-l} \tilde{M}(t,l+u)\frac{r(l+u,l)}{\lambda(t,l)} B(l+u)\kappa \lambda(t,l)du\bigg]...\\
    &...+ \int_0^{t-l}\int_0^u\tilde{M}(t,l+k)\frac{r(l+k,l)}{\lambda(l+u,l)}B(l+k)\kappa \lambda(l+u,l)g(u,l) dk du
\end{align}
and then rearranged to
\begin{equation}
     \tilde{M}(t,l) = \int_0^{t-l} \tilde{M}(t,l+u)r(l+u,l)B(l+u)\kappa (1-G(u,l))du
\end{equation}
Defining $\tilde{S}$ as the analogue to $S$, by
\begin{equation}
    \tilde{S}(t,l,u) = \bigg[\int_0^{u}\tilde{M}(t,l+k)B(l+k)\kappa r(l+k,l)dk\bigg]^2
\end{equation}
and using $\bar{G} = 1-G$, the first of these equations can be written more succinctly as
\begin{equation}
\label{eq:meaneqn1}
    \tilde{M}(t,l) = \bar{G}(t-l,l) \tilde{S}(t,l,t-l)^{0.5} + \int_0^{t-l}\tilde{S}(t,l,u)^{0.5}g(u,l)du
\end{equation}
The equation for $\tilde{V}(t,l)$ can be calculated in a similar way. The only changes to the derivation are that the first term in Supplementary Equation \ref{eq:cumincdifference} is discarded to account for the discarded $s$ in the pgf, and that when adding the mean to move from $W$ to $X$ (in analogue to Supplementary Equation \ref{eq:meanadd}), one no longer needs to add the $\bar{G}(t-l,l)$ term. Thus,
\begin{align}
    \tilde{V}(t,l) &=\bar{G}(t-l,l)\tilde{S}(t,l,t-l) + \int_0^{t-l}\tilde{S}(t,l,u)g(u,l)du+ \int_0^{t-l}\chi^{\tilde{V}}(t,l,k)\bar{G}(k,l)dk - \tilde{M}(t,l)^2
\end{align}
Now, the proof of overdispersion can begin. Firstly, it is helpful to bound $\tilde{M}(t,l)$ above, which can be done as follows. Squaring Supplementary Equation \ref{eq:meaneqn1} shows that
\begin{align}
  \label{eq:overdispersion1} \tilde{M}(t,l)^2 &= \bar{G}(t-l,l)^2\tilde{S}(t,l,t-l) +2\bar{G}(t-l,l) \tilde{S}(t,l,t-l)^{0.5}\int_0^{t-l}\tilde{S}(t,l,u)^{0.5}g(u,l)du+  \bigg[\int_0^{t-l}\tilde{S}(t,l,u)^{0.5}g(u,l)du\bigg]^2
\end{align}
Now, using the Cauchy-Schwarz inequality, we see that
\begin{align}
    \bigg[\int_0^{t-l}\tilde{S}(t,l,u)^{0.5}g(u,l)du\bigg]^2 &= \bigg[\int_0^{t-l}(\tilde{S}(t,l,u)g(u,l))^{0.5} (g(u,l))^{0.5}du\bigg]^2 \\
    &\leq \bigg[\int_0^{t-l}\tilde{S}(t,l,u)g(u,l)du\bigg]\bigg[\int_0^{t-l}g(u)du\bigg] \\
    &\leq (1-\bar{G}(t-l,l))\bigg[\int_0^{t-l}\tilde{S}(t,l,u)g(u,l)du\bigg]
\end{align}
Suppose that $\bar{G}(t-l,l) \neq 1$. Then, using
\begin{equation}
    1 = \frac{1}{1-\bar{G}(t-l,l)} - \frac{\bar{G}(t-l,l)}{1-\bar{G}(t-l,l)}
\end{equation}
to split the final term in Supplementary Equation \ref{eq:overdispersion1}, we find
\begin{align}
    \nonumber \tilde{M}(t,l)^2 &\leq \bar{G}(t-l,l)^2\tilde{S}(t,l,t-l) +2\bar{G}(t-l,l)\tilde{S}(t,l,t-l)^{0.5}\int_0^{t-l}\tilde{S}(t,l,u)^{0.5}g(u,l)du...\\
    &- \frac{\bar{G}(t-l,l)}{1-\bar{G}(t-l,l)} \bigg[\int_0^{t-l}\tilde{S}(t,l,u)^{0.5}g(u,l)du\bigg]^2 + \bigg[\int_0^{t-l}\tilde{S}(t,l,u)g(u,l)du\bigg]
\end{align}
To facilitate the remainder of this proof, it is helpful to define
\begin{equation}
    Q(t,l) := \int_0^{t-l}\tilde{S}(t,l,u)^{0.5}g(u,l)du
\end{equation}
Note that $Q(t,l) \geq 0$ as $\tilde{S}$ and $g$ are non-negative. Moreover, for fixed $t$ and $l$, the function $\tilde{S}(t,l,u)^{0.5}$ is non-decreasing in $u$ and hence
\begin{equation}
      Q(t,l) \leq \int_0^{t-l}\tilde{S}(t,l,t-l)^{0.5}g(u,l)du = \tilde{S}(t,l,t-l)^{0.5}(1-\bar{G}(t-l,l))
\end{equation}
Consider the function
\begin{equation}
    f(Q) = 2\bar{G}(t-l,l) \tilde{S}(t,l,t-l)^{0.5}Q- \frac{\bar{G}(t-l,l)}{1-\bar{G}(t-l,l)}Q^2
\end{equation}
for $Q \in [0,\tilde{S}(t,l,t-l)^{0.5}(1-\bar{G}(t-l,l))]$. $f$ is a quadratic, and has a single turning point at
\begin{equation}
    f'(Q) = 0 \Rightarrow Q = \tilde{S}(t,l,t-l)^{0.5}(1-\bar{G}(t-l,l))
\end{equation}
This is an endpoint of the domain of $Q$ and hence the maximum value of $f(Q)$ must occur one of the endpoints. $f(0) = 0$ and
\begin{align}
    f\bigg(\tilde{S}(t,l,t-l)^{0.5}(1-\bar{G}(t-l,l))\bigg) = \bar{G}(t-l,l)(1-\bar{G}(t-l,l))\tilde{S}(t,l,t-l)
\end{align}
This is non-negative, and hence the maximal value of $f(Q)$. 
\\
\\
\noindent
This can be put into the equation for $\tilde{M}(t,l)^2$ to give
\begin{align}
  \nonumber  \tilde{M}(t,l)^2 &\leq \bar{G}(t-l,l)^2\tilde{S}(t,l,t-l) +\bar{G}(t-l,l)(1-\bar{G}(t-l,l))\tilde{S}(t,l,t-l) + \bigg[\int_0^{t-l}S(t,l,u)g(u,l)du\bigg]\\
    &= \bar{G}(t-l,l)\tilde{S}(t,l,t-l) + \bigg[\int_0^{t-l}S(t,l,u)g(u,l)du\bigg]
\end{align}
Both the terms on the right hand side appear in the equation for $\tilde{V}$, and hence, substituting this result in shows that
\begin{align}
     \tilde{V}(t,l) &\geq  \int_0^{t-l}\chi^{\tilde{V}}(t,l,k)\bar{G}(k,l)dk 
\end{align}
As this holds for all $\bar{G}(t-l,l) < 1$, it must also (under relevant continuity assumptions) hold for $\bar{G}(t-l,l) = 1$, and hence in all cases. Now,
\begin{equation}
    \chi^{\tilde{V}}(t,l,k) = \tilde{V}(t,l+k)B(l+k)r(l+k,l) + \tilde{M}(t,l+k)^2\mathbb{E}(Y(l+k,l)^2)r(l+k,l)
\end{equation}
As $Y \geq 1$ by definition, one has
\begin{equation}
    B(l+k) = \mathbb{E}(Y(l+k,l)) \leq \mathbb{E}(Y(l+k,l)^2)
\end{equation}
and hence
\begin{equation}
     \chi^{\tilde{V}}(t,l,k)\leq  \bigg[\tilde{V}(t,l+k) + \tilde{M}(t,l+k)^2\bigg]B(l+k)r(l+k,l) = \mathbb{E}(\tilde{Z}(t,l+k)^2)B(l+k)r(l+k,l)
\end{equation}
Finally, as $\tilde{Z}(t,l+k) \geq 0$ and is integer-valued, one has $\tilde{Z}(t,l+k)^2 \geq \tilde{Z}(t,l+k)$ and hence
\begin{equation}
     \chi^{\tilde{V}}(t,l,k)\leq \tilde{M}(t,l+k)B(l+k)r(l+k,l)
\end{equation}
Thus,
\begin{align}
\label{eq:overdispersionfinal}
     \tilde{V}(t,l) &\geq  \int_0^{t-l}\tilde{M}(t,l+k)B(l+k)r(l+k,l)\bar{G}(k,l)dk = \tilde{M}(t,l+k)
\end{align}
which proves weak overdispersion.
\\
\\
\noindent
To prove strict overdispersion, note that, for Supplementary Equation \ref{eq:overdispersionfinal} to hold to equality, it is necessary that all the inequalities used hold to equality. Thus, in particular, it is necessary that
\begin{equation}
    \int_0^{t-l}\tilde{M}(t,l+k)B(l+k)r(l+k,l)\bar{G}(k,l)dk = \int_0^{t-l}\mathds{E}(\tilde{Z}(t,l+k)^2)B(l+k)r(l+k,l)\bar{G}(k,l)dk
\end{equation}
and hence, as $B(l+k) \geq 1$,
\begin{equation}
    r(l+k,l)\bar{G}(k,l) \geq 0 \Rightarrow\mathds{E}(\tilde{Z}(t,l+k)^2) = \tilde{M}(t,l+k)
\end{equation}
This means that
\begin{equation}
    r(l+k,l)\bar{G}(k,l) \geq 0 \Rightarrow\mathds{E}(\tilde{Z}(t,l+k)(\tilde{Z}(t,l+k)-1)) = 0
\end{equation}
and hence, as $\tilde{Z}(t,l+k)(\tilde{Z}(t,l+k)-1)$ is a non-negative integer, this means that
\begin{equation}
    r(l+k,l)\bar{G}(k,l) \geq 0 \Rightarrow\tilde{Z}(t,l+k)(\tilde{Z}(t,l+k)-1) = 0
\end{equation}
almost surely. We now show that if $\mathbb{P}(\tilde{Z}(t,l) = 1) > 0$, then $\mathbb{P}(\tilde{Z}(t,l) > 1) > 0$. This can be done as follows. 

Define the set $\mathcal{S}$ to be the possible times at which the initial individual can cause a secondary infection which in turn starts an epidemic that can have non-zero prevalence at time $t$. Then,
\begin{equation}
    \mathcal{S} = \bigg\{ u \in (l,t-l) : r(l+u,l) > 0, \quad \bar{G}(u,l) > 0 \quad \text{and}\quad  \mathbb{P}(Z(t,l+u) > 0) > 0\bigg\}
\end{equation}
Note the use of $Z$ rather than $\tilde{Z}$. The first two conditions ensures that the likelihood of the initial individual causing an infection at time $u$ is non-zero (as it must have non-zero rate here, and also a non-zero probability of still being infectious). The third condition ensures that the probability of this secondary infection's infection tree still containing at least one infectious individual at time $t$ is non-zero. It is necessary that
\begin{equation}
    \int_{\mathcal{S}}r(l+u,l)\bar{G}(u,l)\mathbb{P}(Z(t,l+u) > 0) du > 0
\end{equation}
as otherwise, $\tilde{Z}(t,l) = 0$ (as this integral sums over all possible epidemics that lead to $\tilde{Z}(t,l) > 0$). Define
\begin{equation}
    \mathcal{S}(x) := \mathcal{S}\cap(l,x)
\end{equation}
and the function
\begin{equation}
    f(x) = \int_{\mathcal{S}(x)}r(l+u,l)\bar{G}(u,l)\mathbb{P}(Z(t,l+u) > 0) du 
\end{equation}
Then, $f$ must be continuous, and so there exists some $y \in(0, t-l)$ such that
\begin{equation}
    0 < f(y) < f(t-l) =   \int_{\mathcal{S}}r(l+u,l)\bar{G}(u,l)\mathbb{P}(Z(t,l+u) > 0) du
\end{equation}
Thus, there is a non-zero probability of an individual being infected in $(l,l+y)$ causing an epidemic that has non-zero prevalence at time $t$ and, similarly, a non-zero probability of an individual being infected in $(l+y,t)$ causing an epidemic that has non-zero prevalence at time $t$. Thus, as the infections processes have independent increments and as the initial individual causing an infection in $(l+y,t)$ implies that it must have been infectious for the whole interval $(l,l+y)$, there is a non-zero probability of two such individuals being infected: one in $(l,l+y)$ and one in $(l+y,t)$. Hence
\begin{equation}
    \mathbb{P}(\tilde{Z}(t,l) =1) > 0 \Rightarrow \mathbb{P}(\tilde{Z}(t,l) > 1) > 0
\end{equation}
as required. Thus, 
\begin{equation}
    r(l+k,l)\bar{G}(k,l) \geq 0 \Rightarrow \tilde{Z}(t,l+k) = 0
\end{equation}
and so
\begin{equation}
    \mathds{E}(\tilde{Z}(t,l+k))r(l+k,l)\bar{G}(k,l) = 0\quad \forall k
\end{equation}
Thus, we have strict overdispersion, $\tilde{V}(t,l) > \tilde{M}(t,l)$, provided that the epidemic is expanded at time $t$, as required.

\subsection{Comparison to a Poisson case}
Consider comparing the variance Supplementary Equation \ref{eq:Vareq} with the variance of an epidemic where infection events are always of size 1 (that is, where the counting process of infections, $N^*(t,l)$ is a Poisson case, meaning $B^*(t) = 1$). Asterisks will be used to denote the quantities relating to this Poisson epidemic.

 Suppose that the infectious period is the same in both cases (so $G = G^*$ and $\nu = \nu^*$). To ensure a fair comparison, it is also assumed that the mean number of cases is the same in both cases with $M(t,l) = M^*(t,l)$. By examining the Supplementary Equation \ref{eq:Meaneq} for the mean, and absorbing $\kappa$ into $\rho$ in both cases, one can see
\begin{equation}
    B(l+u)\rho(l+u) = \rho^*(l+u).
\end{equation}
 The variance Supplementary Equation \ref{eq:Vareq} can now be examined. Firstly, note that
\begin{equation}
    \int_0^{t-l}M(t,l+u)B(l+u)r(l+u,l)du = \int_0^{t-l}M^*(t,l+u)r^*(l+u,l)du,
\end{equation}
using the result above and the fact that $M(t,l+u) = M^*(t,l+u)$. Similarly,
\begin{equation}
    S(t,l,u)  = S^*(t,l,u).
\end{equation}
Thus,
\begin{align}
    V(t,l) - V^*(t,l) &= \int_0^{t-l}(\chi^V(t,l,k) - \chi^{V^*}(t,l,k))\bar{G}(k,l)dk\\
    &= \int_0^{t-l}\bigg(V(t,l+k) - V^*(t,l+k)\bigg)B(l+k)r(l+k,l)\bar{G}(k,l)dk...\\
    &+ \int_0^{t-l}\bigg(\mathbb{E}(Y(l+k)^2)-1\bigg)M(t,l+k)^2r(l+k,l)\bar{G}(k,l)dk
\end{align}
By defining $\Delta^V(t,l) :=  V(t,l) - V^*(t,l)$, one can see that this is a renewal equation
\begin{equation}
    \Delta^V(t,l) = \int_0^{t-l}\bigg(\mathbb{E}(Y(l+k)^2)-1\bigg)M(t,l+k)^2r(l+k,l)\bar{G}(k,l)dk +  \int_0^{t-l}\Delta^V(t,l+k)B(l+k)r(l+k,l)\bar{G}(k,l)dk.
\end{equation}
An important property of this renewal equation is that the part that is independent of $\Delta^V$ on the right hand side grows. That is,
\begin{equation}
     \Delta^V(t,l) \geq \int_0^{t-l}\bigg(\mathbb{E}(Y(l+k)^2)-1\bigg)M(t,l+k)^2r(l+k,l)\bar{G}(k,l)dk.
\end{equation}
Thus, even though these two epidemics give the same mean, the difference in their variances is proportional to the square of this mean. This means that models fitted to a Poisson process framework, even without exponential infectious periods, will substantially underestimate the variance of the number of cases (recalling that $\mathbb{E}(Y(l+k)^2) > 1$ in the non-Poisson case).

\subsection{Large time solutions to the variance equation}
To further understand the variance, we consider large time approximate solutions to the variance equation. Note that the level of rigour in this note is lower than the rest of our derivations as the results are derived for illustrative purposes. 

It shall be assumed throughout this note that $\kappa$ has been absorbed into $\rho$. Moreover, to enable explicit asymptotic solutions to be found, it shall be assumed that $\rho$, $B$ and $\mathbb{E}(Y^2)$ are constants and that $g = g(t)$. Therefore all individuals behave identically (in distribution), irrespective of the time at which they were infected. Moreover, it means that $r(l+k,l) = r(k)$,  as the rate of infection depends only on the time since the individual has been infected

Under these assumptions, the mean $M(t,l) = M(t-l)$ and the variance $V(t,l) = V(t-l)$ are functions of $t-l$ only. This property will be used when forming the heuristics used in this note.

The final assumption is that $\bar{G}(t)$ has a finite support - that is, $\bar{G}(t) = 0$ for sufficiently large $t$. This is not strictly necessary, but simplifies the analysis.

Then, for $t > > l$, the mean and variance equations become
\begin{equation}
\label{eq:MeanSimple}
    M(t,l) = \int_0^{t-l}M(t,l+u)B\rho\nu(u)\bar{G}(u)du
\end{equation}
and
\begin{equation}
    V(t,l) = \int_0^{t-l}S(t,l,u)g(u,l) + \int_0^{t-l}\chi^V(t,l,k)\bar{G}(k)dk - M(t,l)^2.
\end{equation}
Motivated by the exponential growth of epidemics without susceptible depletion, consider the heuristic
\begin{equation}
    M(t,l) = e^{\gamma(t-l)}
\end{equation}
for some growth rate $\gamma$ (note that in Supplementary Equation \ref{eq:MeanSimple}, scaling $M$ by a constant does not affect the solution). Then, Supplementary Equation \ref{eq:MeanSimple} becomes
\begin{equation}
    e^{\gamma(t-l)} = e^{\gamma (t-l)}\int_0^{t-l}e^{-\gamma u}B\rho\nu(u)\bar{G}(u)du.
\end{equation}
Now, assuming that $t - l >>1$, as the integrand has finite support,
\begin{equation}
    e^{\gamma(t-l)} = e^{\gamma (t-l)}\int_0^{\infty}e^{-\gamma u}B\rho\nu(u)\bar{G}(u)du = e^{\gamma(t-l)}H(\gamma),
\end{equation}
where $H(\gamma)$ is a monotonically decreasing function such that $H(-\infty) = \infty$ and $H(\infty) = 0$. It is necessary that
\begin{equation}
    H(\gamma) = 1
\end{equation}
and, by the above notes on $H$, there is a unique value for $\gamma$ (independent of $l$) such that this holds. We shall henceforth assume that $\gamma$ is equal to this value.

Note that (by considering the case $\gamma = 0$)
\begin{equation}
    \gamma >0 \Leftrightarrow \int_0^{\infty}B\rho\nu(u)\bar{G}(u)du > 1
\end{equation}
and so the epidemic grows if and only if the expected number of cases caused by an individual is greater than 1, as expected.

The variance equation can now be considered. Note that
\begin{equation}
    S(t,l,u) = \bigg[\int_0^uM(t,l+k)Br(k)dk\bigg]^2 = e^{2\gamma(t-l)}\bigg[\int_0^ue^{-\gamma k}Br(k)dk\bigg]^2.
\end{equation}
Hence, the equation for the variance becomes
\begin{align}
    \nonumber V(t,l) &= e^{2\gamma(t-l)}\int_0^{t-l}\bigg[\int_0^ue^{-\gamma k}Br(k)dk\bigg]^2g(u)du + \int_0^{t-l}V(t,l+k)Br(k)\bar{G}(k)dk...\\
     &+e^{2\gamma(t-l)}\int_0^{t-l}e^{-2\gamma k}\mathbb{E}(Y^2)r(k)\bar{G}(k)dk - e^{2\gamma(t-l)}.
\end{align}
Note the $\chi^V$ term has been split into the two single integrals with integration variable $k$. This equation motivates a heuristic
\begin{equation}
    V(t,l) = Ce^{2\gamma(t-l)},
\end{equation}
which, again using the fact that the integrands have finite support, results in
\begin{equation}
    C = \frac{\int_0^{\infty}\bigg[\int_0^ue^{-\gamma k}Br(k)dk\bigg]^2g(u)du + \int_0^{\infty}e^{-2\gamma k}\mathbb{E}(Y^2)r(k)\bar{G}(k)dk - 1}{1 - \int_0^{\infty}e^{-2\gamma k}Br(k)\bar{G}(k)dk}.
\end{equation}
Note that
\begin{align}
    \int_0^{\infty}\bigg[\int_0^ue^{-\gamma k}Br(k)dk\bigg]^2g(u)du &> \int_0^{\infty}\bigg[\int_0^ue^{-\gamma k}Br(k)\bar{G}(k)dk\bigg]^2g(u)du\\
    &= \int_0^{\infty}g(u)du\\
    &=1.
\end{align}
and hence the numerator is strictly positive.

Moreover, suppose that $\gamma > 0$.  Then, note that
\begin{equation}
    \int_0^{\infty}e^{-2\gamma k}Br(k)\bar{G}(k)dk < \int_0^{\infty}e^{-\gamma k}Br(k)\bar{G}(k)dk = 1
\end{equation}
which means that the denominator (and hence $C$) is strictly positive. 

Note that if $\gamma \leq 0$, this variance approximation is not well-defined (as $C$ is either infinite if $\gamma = 0$ or negative if $\gamma < 0$) and so it is necessary to find another solution. In the $\gamma < 0$ case, $e^{\gamma(t-l)} >> e^{2\gamma(t-l)}$ and a leading-order solution can be found simply from
\begin{equation}
    V = e^{\gamma(t-l)}.
\end{equation}
Thus, according to these approximations, the variance grows with the square of the mean in the $\gamma > 0$ (i.e. growing epidemic) case, while it decreases proportionally to the mean in the $\gamma < 0$ (i.e. shrinking epidemic) case. The $\gamma = 0$ case is the bifurcation point between these two solutions and would require further analysis.

In the growing epidemic case, the equation for $C$ is also informative in characterising the effect of the different model parameters on the variance. In particular, it shows that there is a linear relationship between $\mathds{E}(Y(t)^2)$ and the variance, re-emphasising the point made in the previous subnote that ignoring this parameter can have significant effects on the variance estimate. Moreover, it shows that variance grows rapidly throughout a growing epidemic, remaining proportional to the square of the mean. 
\subsection{Mean and variance for cumulative incidence}
The equations for the mean and prevalence of the cumulative incidence of the epidemic can be derived almost identically, as the two generating functions are very similar. The mean equation gains an term from the additional $s$ being differentiated, which is
\begin{equation}
    \int_0^{t-l}\mathcal{J}_{(l+u)}\bigg(\int_0^u\mathcal{Y}_{(l+k,l)(1)}\frac{r(l+k,l)}{\lambda(l+u,l)}dk\bigg)g(u,l)du = G(t-l,l)
\end{equation}
and hence, the mean equation becomes (using *s to denote cumulative incidence quantities)
\begin{equation}
    M^*(t,l) = 1+\int_0^{t-l}M^*(t,l+u)\rho(l+u)\nu(u)\bar{G}(u,l)du
\end{equation}
Now, the only difference in the equation for $W$ in the case of cumulative incidence is that the term Supplementary Equation \ref{eq:cumincdifference} appears in both parts (again due to the extra $s$ term). This can be treated in the same way as $\chi$ in the original derivation and so
\begin{align}
    W^*(t,l) &= 2\int_0^{t-l}\kappa M^*(t,l+u)B(l+u)r(l+u,l)\bar{G}(u,l)du +\bar{G}(t-l,l)\tilde{S}(t,l,t-l)...\nonumber\\
    &...+ \int_0^{t-l}\tilde{S}(t,l,u)g(u,l)du+ \int_0^{t-l}\chi^*(t,l,k)\bar{G}(k,l)dk
\end{align}
Again, following the previous derivation, one can then arrive at
\begin{align}
       V^*(t,l) &= 1 + 2\int_0^{t-l}\kappa M^*(t,l+u)B(l+u)r(l+u,l)\bar{G}(u,l)du +\bar{G}(t-l,l)\tilde{S}(t,l,t-l)...\nonumber\\
    &...+ \int_0^{t-l}\tilde{S}(t,l,u)g(u,l)du+ \int_0^{t-l}\chi^{V^*}(t,l,k)\bar{G}(k,l)dk - M^*(t,l)^2 
\end{align}
\section{Likelihood functions}
\subsection{Continuous case}
If only the cumulative incidence, $Z(t,l)$, is known at some time $t$, the full epidemic history - in particular, the times at which each individual was infected, and the times at which they stopped being infectious - are unknown. Thus, it is helpful to derive a likelihood function for each possible sequence of these times.
\\
\\
\noindent
Perhaps the most intuitive approach would be to treat the times at which each individual was infected as continuous random variables. However, the resultant pdf is complicated by the fact that multiple infections are likely to happen simultaneously if $\mathbb{E}(Y) > 1$, and will have a significant number of Kronecker delta functions to accommodate this, making it complicated both mathematically and practically.
\\
\\
\noindent
To remedy this, we instead consider three sets of random variables - a vector $\boldsymbol{T}$ of unknown size $n+1$, which contains the times of all the infection events up to time $t$; a vector $\boldsymbol{Y}$ also of size $n+1$, which contains the size of each of these infection events (that is, $y_m$ is the number of individuals that are infected at time $\tau_m$); and a vector $\boldsymbol{D}$ containing the times at which each individual stops being infected. To make the subsequent notation clearer, we shall use a non-rectangular array $\boldsymbol{X}$ in place of $\boldsymbol{D}$, where $X_{ij}$ will be the time at which the jth individual infected at time $T_i$ stops being infected.
\\
\\
\noindent
We will suppose that for each $s>u$ and positive integer $k$
\begin{equation}
    \mathbb{P}(N(s+dt,u) -N(s,u) = k)  = p_k(s,u)dt + o(dt)
\end{equation}
and that
\begin{equation}
    \mathbb{P}(N(s+dt,u) -N(s,u) = 0) = 1 - \sum_{k\geq 1}p_k(s,u)dt + o(dt) = 1-r(s,u)dt + o(dt)
\end{equation}
as the counting process of jumps in $N(s,u)$ is an inhomogeneous Poisson Process of rate $r(s,u)$ (absorbing the $\kappa$ into $r$). We can hence create a likelihood function. Define $\boldsymbol{1}$ to be a vector of 1's, and choose any vectors $\boldsymbol{\tau}$ and $\boldsymbol{d}$ such that each $\tau_i,d_j \in (0,t)$. Define $dt$ to be small enough so that $\tau_i - \tau_j > dt$ for all $i > j$ and so that $|\tau_i - d_j| > dt$ for all $i,j$ (note that, the set where $\tau_i = d_j$ has zero measure and can be ignored). Moreover, choose a positive-integer-valued vector $\boldsymbol{y}$. Then,
\begin{align}
    \nonumber&\mathbb{P}(\boldsymbol{T} \in [\boldsymbol{\tau},\boldsymbol{\tau}+dt\boldsymbol{1}], \boldsymbol{D} \in[\boldsymbol{d},\boldsymbol{d}+dt\boldsymbol{1}],\boldsymbol{Y} = \boldsymbol{y})
    = P\bigg[\bigg(\bigcap_{k=1}^n\{y_k \text{ infections in } [\tau_k,\tau_k+dt]\}\bigg)...\\
    &...\cap \bigg(\bigcap_{k=0}^n \{ \text{no infections in }[\tau_k+dt,\tau_{k+1}]\bigg)\cap\bigg(\bigcap_{i=0}^{n}\bigcap_{j=1}^{y_i}\{L \in [x_{ij}-\tau_i, x_{ij} - \tau_i+dt]\}\bigg)\bigg]
\end{align}
where $\tau_{n+1} := t$ to reduce notation, $x_{ij}$ is the value of $X_{ij}$ in the case $\boldsymbol{D} = \boldsymbol{d}$ and $L$ is a random variable equal in distribution to the infectious period of an individual. Each of the infection events in the above equation occur on disjoint subintervals of $[0,t]$ and so, as all of the processes $N(t,l)$ have independent increments, and each individual behaves independently of each other and their infectious periods, they can be considered separately. We have
\begin{equation}
    \mathbb{P}(y_k \text{ infections in } [\tau_k,\tau_k+dt]) = \sum_{i = 0}^{k-1}\sum_{j=0}^{y_i} \mathds{1}_{\{x_{ij} < \tau_k\}} p_{y_k}(\tau_k,\tau_i)dt + o(dt)
\end{equation}
Here, the $o(dt)$ term contains three components that can be linearised out of the model - the probability that multiple different individuals contribute to the $y_k$ cases (this is $O(dt^2)$); the probabilities of individuals infecting no one in this interval (these are independently $1-O(dt)$ and hence the $O(dt)$ contribution can be ignored when these probabilities are multiplied together); and the $o(dt)$ terms from the equations defining $p_k$. 
\\
\\
\noindent
As the counting process of jumps in $N(s,u)$ is an inhomogeneous Poisson Process, and it is only ``active'' for individual ${ij}$ up to time $x_{ij}$,
\begin{equation}
    \mathbb{P}(\text{no infections in }[\tau_k+dt,\tau_{k+1}]) = \prod_{i=0}^k\prod_{j=1}^{y_i}\exp\bigg(-\int_{\min(x_{ij},\tau_k)}^{\min(x_{ij},\tau_{k+1})}r(u,\tau_i)du\bigg) + O(dt)
\end{equation}
where here, the $O(dt)$ term contains the integral between $\tau_k$ and $\tau_k+dt$ of each of the integrands. Taking the products inside the exponential as sums, the various ``no infection'' terms can be combined together to give 
\begin{equation}
    P\bigg(\bigcap_{k=0}^n \{ \text{no infections in }[\tau_k+dt,\tau_{k+1}]\}\bigg) = \exp\bigg(-\sum_{i=0}^n\sum_{j=0}^{y_i}\int_{\tau_i}^{\min(t,x_{ij})}r(u,\tau_i)du\bigg)
\end{equation}
Finally, the infectious period terms can be simply calculated from the pdf, $g$, of $L$ as
\begin{equation}
    \mathbb{P}(L \in [x_{ij}-\tau_i, x_{ij} - \tau_i+dt]) = g(x_{ij} - \tau_i,\tau_i)dt + o(dt)
\end{equation}
Hence, combining all the relevant terms,
\begin{align}
       \nonumber&\mathbb{P}(T \in [\boldsymbol{\tau},\boldsymbol{\tau}+dt\boldsymbol{1}], \boldsymbol{D} \in[\boldsymbol{d},\boldsymbol{d}+dt\boldsymbol{1}],\boldsymbol{Y} = \boldsymbol{y}) = o(dt^{n+Z(t,l)}) + \\
       &\prod_{k=1}^n\bigg[\bigg(\prod_{j=1}^{y_k}g(x_{kj} - \tau_k,\tau_k)\bigg)\bigg(\sum_{i = 0}^{k-1}\sum_{j=0}^{y_i} \mathds{1}_{\{X_{ij} < \tau_k\}} p_{y_k}(\tau_k,\tau_i)\bigg)\bigg] \exp\bigg(-\sum_{i=0}^n\sum_{j=0}^{y_i}\int_{\tau_i}^{\min(t,X_{ij})}r(u,\tau_i)du\bigg)(dt)^{n+Z(t,l)} 
\end{align}
and thus, taking $dt \to 0$ gives a likelihood function of
\begin{equation}
    L(\boldsymbol{\tau},\boldsymbol{y},\boldsymbol{d}) = \prod_{k=1}^n\bigg[\bigg(\prod_{j=1}^{y_k}g(x_{kj} - \tau_k,\tau_k)\bigg)\bigg(\sum_{i = 0}^{k-1}\sum_{j=0}^{y_i} \mathds{1}_{\{x_{ij} < \tau_k\}} p_{y_k}(\tau_k,\tau_i)\bigg)\bigg] \exp\bigg(-\sum_{i=0}^n\sum_{j=0}^{y_i}\int_{\tau_i}^{\min(t,x_{ij})}r(u,\tau_i)du\bigg)
\end{equation}
It is simple to substitute in the two examples that have been previously considered. In both cases, $r(a,b) = \rho(a)\nu(a-b)$. In the Poisson case, one has $p_1(a,b) = \rho(a)\nu(a-b)$ and $p_k(a,b) = 0$ for $k > 1$. In the Negative Binomial case, the values of $p_k$ are given by
\begin{align}
   p_k(a,b)dt &= \lim_{t \to 0}\bigg(\frac{\mathbb{P}(T \in [\boldsymbol{\tau},\boldsymbol{\tau}+dt\boldsymbol{1}], \boldsymbol{D} \in[\boldsymbol{d},\boldsymbol{d}+dt\boldsymbol{1}],\boldsymbol{Y} = \boldsymbol{y})}{dt^{n+Z(t,l)} }\bigg)\\
   &=\mathbb{P}(J_M(a+dt,b) - J_M(a,b)= 1)\mathbb{P}(Y = k) \\
   &=  \rho(a)\nu(b-a)\bigg( \frac{(1-p)^k}{-k\ln(p)}\bigg)
\end{align}

\subsection{Special case (Poisson)}
In the Poisson case, $A_{k,i}$ is Poisson distributed with mean $\rho(k)\nu(k-i)$. Hence,
\begin{equation}
    \mathcal{A}_k(\boldsymbol{b},\boldsymbol{y},\boldsymbol{d}) \sim \text{Poi}\bigg(\rho(k)\sum_{i=0}^{k-1}\nu(k-i)\sum_{j=1}^{y_i} \mathds{1}_{\{x_{ij} \leq k\}}  \bigg) := \text{Poi}(\mu_k)
\end{equation}
and so, the more computationally useful log-likelihood is
\begin{equation}
     \ell(\boldsymbol{\tau},\boldsymbol{y},\boldsymbol{D}) =\sum_{k=1}^n(\mu_k\log(y_k) - \mu_k - \log(y_k!)) + \sum_{i=1}^n\sum_{j=1}^{y_i} \log(g(x_{ij}-\tau_i,\tau_i))
\end{equation}
\subsection{Special case (Negative Binomial)}
In the Negative Binomial case,
\begin{equation}
    A_{k,i} =_D \sum_{j=1}^{N}Y_j
\end{equation}
where the $Y_j$ are iid logarithmic random variables with a pmf given by Supplementary Equation \ref{eq:logpdf} that is independent of the properties of the individual, and $N$ is Poisson distributed with mean $\rho(k)\nu(k-i)$. Thus,
\begin{equation}
    \mathcal{A} \sim \mathcal{NB}(\phi \mu_k,p)
\end{equation}
where, as before, $p = \frac{\phi}{1+\phi}$ and $\mu_k$ is defined in the previous note. Hence, as
\begin{equation}
    \log\bigg[ P\bigg(NB(a,p) = k\bigg) \bigg] = \sum_{j = 0}^{k-1}\log(a+j)  + k \log(1-p) + a\log(p) - \log(k!)
\end{equation}
we have
\begin{align}
   \ell(\boldsymbol{\tau},\boldsymbol{y},\boldsymbol{D}) = \sum_{k=0}^n\bigg[\log(\phi \mu_k+j)  + y_k \log\bigg(\frac{1}{1+\phi}\bigg) + \phi \mu_k\log\bigg(\frac{\phi}{1+\phi}\bigg) - \log(y_k!)\bigg] + \sum_{i=1}^n\sum_{j=1}^{y_i} \log(g(x_{ij}-\tau_i,\tau_i))
\end{align}
\subsection{Approximating the likelihood}
It is difficult to simulate from the likelihoods when the infectious periods of the individuals are unknown because often, $Z(t,l) >> t$ (whereas the other unknowns, $\boldsymbol{\tau}$ and $\boldsymbol{y}$ have only $n \sim t$ parameters). To remedy this, we use an approximation - given an estimate of the function $g$, we simulate
\begin{equation}
    D_i = L_i + \tau_i \quad \text{where $L_i \sim g$}
\end{equation}
For some $\boldsymbol{D}$, the observed epidemic may be impossible (e.g. if, $D_0 < b_1$, where $b_1$ is the time that the first infection event occurs). Thus, it necessary to impose a feasibility condition. Many such conditions are possible, but we use a simple condition by defining
\begin{equation}
    L_i^* := \max(L_i,\tau_{i+1}-\tau_i)
\end{equation}
and then define
\begin{equation}
    D_i^* := \tau_i + L_i^*
\end{equation}
Given these values, we can then create an approximation, $\ell^*$ to be
\begin{equation}
    \ell^*(\boldsymbol{\tau},\boldsymbol{y}) \sim \ell(\boldsymbol{\tau},\boldsymbol{y},\boldsymbol{D}^*)
\end{equation}
This clearly creates a non-deterministic likelihood as it is dependent on a set of random variables. However, from our simulations, it appears that $\ell^*$ has a small variance, and so this extra randomness does not significantly affect our calculations.
\section{Assessing future variance during an epidemic}
Many of the equations presented thus far have been concerned with properties of an epidemic started from a single case at a fixed deterministic time. However, it is crucial to be able to calculate the risk from any time during the epidemic, and such a derivation is presented in this note. This derivation is more algebraically involved than the other work in this paper, and so to reduce its length, it will be assumed that $N(t,l)$ is an inhomogeneous Poisson Process, and that $L = \infty$ for each individual. This means that $\boldsymbol{y}$ and $\boldsymbol{D}$ can be ignored when considering the likelihood.
\subsection{Derivation}
Suppose that the prevalence (or, equivalently in this case, cumulative incidence), $Z(t,l) = n+1$, is known at some point in an epidemic, but that the times at which these infections happened, $B_i$, are unknown. Note that the notation $B_i$ rather than $T_i$ is used in this note, because these times are now an exact analogue of birth times in a birth-death process. The condition of $n+1$ rather than $n$ has been chosen as this means that there have been $n$ new infections and will make the following derivation notationally simpler.
\\
\\
\noindent
Note that the infection time of the initial individual, $B_0$ is known to be equal to $l$, but it will be treated identically to the other times to reduce notation. Its marginal pdf is $f_{B_0}(b) = \delta(b-l)$. Following the previous note, the pdf $f_{\boldsymbol{B}}(\boldsymbol{b})$ of the infection times is
\begin{equation}
    f_{\boldsymbol{B}}(\boldsymbol{b}) =\frac{1}{\mathbb{P}(Z(t,l) = n)} \prod_{i=1}^n\bigg(\rho(b_i)\sum_{j=0}^{i-1}\nu(b_i-b_j)\bigg)\exp\bigg[-\sum_{i=0}^n \int_0^{t-b_i} \rho(s+l)\nu(s)ds\bigg]
\end{equation}
Now, one can write
\begin{equation}
    Z(t+s,l) = \sum_{i=0}^nZ_i^*(t+s,B_i)
\end{equation}
where $Z_i^*(t+s,B_i)$ counts the infection tree started at the individual infection at $b_i$, considering only those infections that occurred after time $t$ (that is, if this individual infects someone at time $a < t$, the infections of this second individual will \emph{not} be counted, even if they occur after time $t$).
\\
\\
\noindent
This can be rewritten as
\begin{equation}
    Z(t+s,l) = \int_{b= 0}^t\sum_{i=0}^nZ_i^*(t+s,b)\mathds{1}_{\{B_i = b\}}
\end{equation}
where here, $\mathds{1}$ is the indicator function. Hence,
\begin{align}
    \text{var}(Z(t+s,l)) &= \text{var}\bigg(\int_{b=0}^t\sum_{i=0}^nZ_i^*(t+s,b)\mathds{1}_{\{B_i = b\}}\bigg)\\
    \nonumber&= \int_{b=0}^t\sum_{i=0}^n\text{var}(Z_i^*(t+s,b)\mathds{1}_{\{B_i = b\}})...\\
    &...+ \int_{b=0}^t\int_{c=0}^t\sum_{i=0}^n\sum_{j=0}^n\text{cov}\bigg(Z_i^*(t+s,b)\mathds{1}_{\{B_i = b\}},Z_j^*(t+s,b)\mathds{1}_{\{B_j = c\}}\bigg)(\mathds{1}_{\{(b,i)\neq (c,j)\}})
\end{align}
The first term in this equation can be expanded as
\begin{align}
    \text{var}(Z_i^*(t+s,b)\mathds{1}_{\{B_i = b\}}) &= \mathbb{E}(Z_i^*(t+s,b)^2\mathds{1}_{\{B_i = b\}}^2) -  \mathbb{E}(Z_i^*(t+s,b)\mathds{1}_{\{B_i = b\}})^2\\
    &= \mathbb{E}(Z_i^*(t+s,b)^2)\mathbb{E}(\mathds{1}_{\{B_i = b\}}) -  \mathbb{E}(Z_i^*(t+s,b))^2\mathbb{E}(\mathds{1}_{\{B_i = b\}})^2
\end{align}
Note that $\mathbb{E}(\mathds{1}_{\{B_i = b\}})^2 = O(db^2)$ and hence this term has zero measure (as it is only integrated over one dimension). This leaves
\begin{equation}
    \text{var}(Z_i^*(t+s,b)\mathds{1}_{\{B_i = b\}}) = \mathbb{E}(Z_i^*(t+s,b)^2)f_{B_i}(b)db
\end{equation}
where $f_{B_i}(b)$ is the marginal pdf of $B_i$. 
\\
\\
\noindent
The second term can also be expanded - note that, by the independence of the $Z^*$ terms, for $i \neq j$
\begin{equation}
\label{eq:covbreak}
    \text{cov}\bigg(Z_i^*(t+s,b)\mathds{1}_{\{B_i = b\}},Z_j^*(t+s,b)\mathds{1}_{\{B_j = c\}}\bigg) = \mathbb{E}(Z_i^*(t+s,b))\mathbb{E}(Z_j^*(t+s,c))\text{cov}(\mathds{1}_{\{B_i = b\}},\mathds{1}_{\{B_j = c\}})
\end{equation}
Moreover, if $i = j$, then one has $b\neq c$ and hence
\begin{equation}
    \mathds{1}_{\{B_i = b\}}\mathds{1}_{\{B_j = c\}} = \mathds{1}_{\{B_i =b, B_i = c\}}= 0
\end{equation}
which means
\begin{align}
    \text{cov}\bigg(Z_i^*(t+s,b)\mathds{1}_{\{B_i = b\}},Z_j^*(t+s,b)\mathds{1}_{\{B_j = c\}}\bigg)  &=-\mathbb{E}(Z_i^*(t+s,b))\mathbb{E}(Z_j^*(t+s,c))\mathbb{E}(\mathds{1}_{\{B_i = b\}})\mathbb{E}(\mathds{1}_{\{B_j = c\}})\\
    &=\mathbb{E}(Z_i^*(t+s,b))\mathbb{E}(Z_j^*(t+s,c))\text{cov}(\mathds{1}_{\{B_i = b\}},\mathds{1}_{\{B_j = c\}})
\end{align}
and hence the Supplementary Equation \ref{eq:covbreak} holds in all cases. Now, one has, for $i \neq j$
\begin{align}
    \text{cov}(\mathds{1}_{\{B_i = b\}},\mathds{1}_{\{B_j = c\}}) &= \mathbb{E}(\mathds{1}_{\{B_i = b\}}\mathds{1}_{\{B_j = c\}}) - \mathbb{E}(\mathds{1}_{\{B_i = b\}})\mathbb{E}(\mathds{1}_{\{B_j = c\}})\\
    &=  \mathbb{E}(\mathds{1}_{\{B_i = b,B_j = c\}}) - f_{B_i}(b)f_{B_j}(c)dbdc\\
    &= (f_{B_i,B_j}(b,c) - f_{B_i}(b)f_{B_j}(c))dbdc
\end{align}
while if $i = j$ and $b \neq c$, this result also holds, following the convention that
\begin{equation}
    f_{B_i,B_i}(b,c) = \delta(b-c)f_{B_i}(b)
\end{equation}
(and hence in this case is zero) where $\delta$ is the Kronecker delta.
\\
\\
\noindent
Thus, in all cases
\begin{equation}
     \text{cov}\bigg(Z_i^*(t+s,b)\mathds{1}_{\{B_i = b\}},Z_j^*(t+s,b)\mathds{1}_{\{B_j = c\}}\bigg)  =\mathbb{E}(Z_i^*(t+s,b))\mathbb{E}(Z_j^*(t+s,c))(f_{B_i,B_j}(b,c) - f_{B_i}(b)f_{B_j}(c))dbdc
\end{equation}
This gives an equation of
\begin{align}
    \nonumber &\text{var}(Z(t+s,l)) =\int_{b=0}^t\sum_{i=0}^n\mathbb{E}(Z_i^*(t+s,b)^2)f_{B_i}(b)db...\\
    \label{eq:varianceZ} &...+ \int_{b=0}^t\int_{c=0}^t\sum_{i=0}^n\sum_{j=0}^n\mathbb{E}(Z_i^*(t+s,b))\mathbb{E}(Z_j^*(t+s,c))(f_{B_i,B_j}(b,c) - f_{B_i}(b)f_{B_j}(c))\mathds{1}_{\{(b,i)\neq (c,j)\}}dbdc
\end{align}
It is more informative to remove the $\mathds{1}_{\{(b,i) \neq (c,j)\}}$ condition. This can be done by calculating
\begin{align}
    &\int_{b=0}^t\int_{c=0}^t\sum_{i=0}^n\sum_{j=0}^n\mathbb{E}(Z_i^*(t+s,b))\mathbb{E}(Z_j^*(t+s,c))\bigg(f_{B_i,B_j}(b,c) - f_{B_i}(b)f_{B_j}(c)\bigg)\mathds{1}_{\{(b,i) = (c,j)\}}dbdc\\
    &=\int_{b=0}^t\int_{c=0}^t\sum_{i=0}^n\mathbb{E}(Z_i^*(t+s,b)\mathbb{E}(Z_i^*(t+s,c))\bigg(\delta(b-c)f_{B_i}(b)- f_{B_i}(b)f_{B_i}(c)\bigg)\mathds{1}_{\{b=c\}}dbdc\\
    &=\int_{b=0}^t\int_{c=0}^t\sum_{i=0}^n\mathbb{E}(Z_i^*(t+s,b)\mathbb{E}(Z_i^*(t+s,c))\bigg(\delta(b-c)f_{B_i}(b)- f_{B_i}(b)f_{B_i}(c)\mathds{1}_{\{b=c\}}\bigg)dbdc\\
       \label{eq:correction} &= \int_{b=0}^t\sum_{i=0}^n\mathbb{E}(Z_i^*(t+s,b))^2f_{B_i}(b)db 
\end{align}
noting that the second term is bounded and contains $\mathds{1}_{\{b=c\}}$ which is non-zero only on a null set of the domain of integration (and hence the integral is zero). Thus, absorbing this correction term into the first term in Supplementary Equation \ref{eq:varianceZ},
\begin{align}
    \nonumber&\text{var}(Z(t+s,l)) =\int_{b=0}^t\sum_{i=0}^n\text{var}(Z_i^*(t+s,b))f_{B_i}(b)db...\\
    \label{eq:var_separ}&...+ \int_{b=0}^t\int_{c=0}^t\sum_{i=0}^n\sum_{j=0}^n\mathbb{E}(Z_i^*(t+s,b))\mathbb{E}(Z_j^*(t+s,c))(f_{B_i,B_j}(b,c) - f_{B_i}(b)f_{B_j}(c))dbdc
\end{align}
The advantage of this formulation is that it allows the contributions to the variance from the infection times $B_i$ before time $t$ and from further infections between times $t$ and $t+s$ to be separated. Indeed, note that if the infection times are known (so that $f_{B_i}(b) = \delta(b-b_i)$), one has
\begin{align}
    \nonumber&\int_{b=0}^t\int_{c=0}^t\sum_{i=0}^n\sum_{j=0}^n\mathbb{E}(Z_i^*(t+s,b))\mathbb{E}(Z_j^*(t+s,c))(f_{B_i,B_j}(b,c) - f_{B_i}(b)f_{B_j}(c))dbdc \\
    &...= \int_{b=0}^t\int_{c=0}^t\sum_{i=0}^n\sum_{j=0}^n\mathbb{E}(Z_i^*(t+s,b))\mathbb{E}(Z_j^*(t+s,c))(\delta(b-b_i)\delta(c-b_j) - \delta(b-b_i)\delta(c-b_j))dbdc \\
    &=0
\end{align}
noting that the definition of
\begin{equation}
     f_{B_i,B_i}(b,c) = f_{B_i}(b)\delta(b-c) =  f_{B_i,B_i}(b,c) = \delta(b-b_i)\delta(b-c) = \delta(b-b_i)\delta(c-b_i)
\end{equation}
is consistent in this case. Thus, the second term in Supplementary Equation \ref{eq:var_separ} is only non-zero when there is uncertainty in the infection times (while, moreover, the first term is only non-zero when there is uncertainty in the infections that occur in the interval $(t,t+s)$, as otherwise $\text{var}(Z_i^*(t+s,b_i)) = 0$).
\\
\\
\noindent
To complete the derivation of the variance equation, it is necessary to derive formulae to calculate the quantities $\text{var}(Z_i^*)$. To enable this, define $M^*(t+s,b_i) := \mathbb{E}(Z_i^*(t+s,b_i))$ and $X^*(t+s,b_i) := \mathbb{E}(Z_i^*(t+s,b_i)^2)$ to be the mean and squared mean of the infection tree started from time $t$ by the ith individual. 
\\
\\
\noindent
These quantities can be calculated directly from the mean and variance, $M$ and $V$, of the ``standard case'' (where a single initial individual is infected at some time $l$), considered in previous notes in this appendix. This is possible as, in the context of renewal processes, the quantities $Z_i^*$ are renewal processes where all but the first individuals are identical, and hence are amenable to similar methodology. Indeed, if one supposes that $\{Z(t+s,t+u)\}_{u \leq s}$ are a set of independent realisations of different ``standard'' epidemics, one has
\begin{equation}
    Z_i^*(t+s,t+u) = \int_{u=0}^sZ(t+s,t+u)\mathds{1}_{\{\text{individual $i$ infects another individual at time $t+u$}\}}
\end{equation}
as the newly infected individuals start new, independent and ``standard'' epidemics. Define
\begin{equation}
    \mathcal{I}_u := \mathds{1}_{\{\text{individual $i$ infects another individual at time $t+u$}\}}
\end{equation}
Hence,
\begin{align}
    M^*(t+s,b_i) &= E\bigg(\int_{u=0}^{s}Z(t+s,t+u)\mathcal{I}_u\bigg)\\
    &= \int_{u=0}^{s}M(t+s,t+u)\rho(t+u)\nu(t-b_i+u)du
\end{align}
Moreover,
\begin{align}
    X^*(t+s,b_i) &=E\bigg(\bigg[\int_{u=0}^{s}Z(t+s,t+u)\mathcal{I}_u\bigg]^2\bigg)\\
    &= E\bigg(\int_{u=0}^{s}\int_{k=0}^{s}Z(t+s,t+u)\mathcal{I}_uZ(t+s,t+k)\mathcal{I}_k\bigg)
\end{align}    
Note that, for $k \neq u$, the quantities $Z(t+s,t+u)$ and $Z(t+s,t+k)$ are independent. Moreover, these quantities are all independent from the indicator terms. Thus, it is helpful to split the integral, giving
\begin{align}    
     W^*(t+s,b_i)
    &= \int_{u=0}^{s}E\bigg(Z(t+s,t+u)^2\mathcal{I}_u\bigg) + \int_{u=0}^s\int_{k=0}^sE\bigg[Z(t+s,t+u)\mathcal{I}_uZ(t+s,t+k)\mathcal{I}_k\bigg]\mathds{1}_{\{u\neq k\}}\\
    &= \int_{u=0}^{s}E\bigg(Z(t+s,t+u)^2\mathcal{I}_u\bigg) + \int_{u=0}^s\int_{k=0}^sM(t+s,t+u)\mathbb{E}(\mathcal{I}_u)M(t+s,t+k)\mathbb{E}(\mathcal{I}_k)\mathds{1}_{\{u\neq k\}}
\end{align}
Now,
\begin{equation}
   \int_{u=0}^s\int_{k=0}^sM(t+s,t+u)\mathbb{E}(\mathcal{I}_u)M(t+s,t+k)\mathbb{E}(\mathcal{I}_k)  = \bigg[\int_{u=0}^s M(t+s,t+u)\mathbb{E}(\mathcal{I}_u)\bigg]^2 = M^*(t+s,b_i)^2
\end{equation}
while
\begin{align}
     \int_{u=0}^s\int_{k=0}^sM(t+s,t+u)\mathbb{E}(\mathcal{I}_u)M(t+s,t+k)\mathbb{E}(\mathcal{I}_k)\mathds{1}_{\{u=k\}} =0
\end{align}
as the integrand is bounded and is non-zero only on a null set of the domain of integration. Hence, one has
\begin{equation}
    \int_{u=0}^s\int_{k=0}^sM(t+s,t+u)\mathbb{E}(\mathcal{I}_u)M(t+s,t+k)\mathbb{E}(\mathcal{I}_k)\mathds{1}_{\{u\neq k} = M^*(t+s,b_i)^2
\end{equation}
Thus,
\begin{align}
    X^*(t+s,b_i) &= \int_{u=0}^{s}E\bigg(Z(t+s,t+u)^2\mathcal{I}_u\bigg) + M^*(t+s,b_i)^2\\
    &= \int_{u=0}^{s}E\bigg(Z(t+s,t+u)^2\bigg)\mathbb{E}(\mathcal{I}_u) + M^*(t+s,b_i)^2\\
    &= \int_{u=0}^{s}(V(t+s,t+u) + M(t+s,t+u)^2)\rho(t+u)\nu(t-b_i+u)du+ M^*(t+s,b_i)^2
\end{align}
Hence, defining $V^*(t+s,b_i) := \text{var}(Z^*(t+s,b_i)) = X^*(t+s,b_i) - M^*(t+s,b_i)^2$, one has
\begin{equation}
    V^*(t+s,b_i) = \int_{u=0}^{s}(V(t+s,t+u) + M(t+s,t+u)^2)\rho(t+u)\nu(t-b_i+u)du
\end{equation}
Hence, one has the final form of the variance equation
\begin{align}
    &\text{var}(Z(t+s,l)) =\int_{b=0}^t\sum_{i=0}^nV^*(t+s,b)f_{B_i}(b)db...\nonumber\\
    &...+ \int_{b=0}^t\int_{c=0}^t\sum_{i=0}^n\sum_{j=0}^nM^*(t+s,b)M^*(t+s,c)(f_{B_i,B_j}(b,c) - f_{B_i}(b)f_{B_j}(c))dbdc
\end{align}
\subsection{Bounding the equation}
Unlike previous formulae, this is an explicit equation and no recursion is required to get the desired results (although recursion is necessary to calculate the $V$ term in $V^*$). However, the infection time pdf makes this a difficult equation to evaluate.
\\
\\
\noindent
However, one can give a simpler upper bound on the variance. Define
\begin{equation}
    \nu_{\text{bound}}(u) := \max_{b_i \in [l,t]}(\nu(t-b_i+u))
\end{equation}
so that
\begin{align}
    M^*(t+s,b_i) \leq  \int_0^{s}M(t+s,t+u)\rho(t+u)\nu_{\text{bound}}(t-b_i+u)du := \mathcal{M}^*(t+s)
\end{align}
and
\begin{equation}
    V^*(t+s,b_i) \leq  \int_{u=0}^{s}(V(t+s,t+u) + M(t+s,t+u)^2) \rho(t+u)\nu_{\text{bound}}(u)du  := \mathcal{V}^*(t+s)
\end{equation}
so that this is now independent of $b_i$. Note that the construction of $\nu_{\text{bound}}(u)$ means that it will still decay for large $u$. Under the assumption that the infection times are roughly deterministic so the second term is zero,
\begin{equation}
     \text{var
     }(Z(t+s,l)) \leq  Z(t,l)\mathcal{V}^*(t+s)
\end{equation}
The covariance term can be added in by noting that
\begin{align}
  \nonumber &\int_{b=0}^t\int_{c=0}^t\sum_{i=0}^n\sum_{j=0}^nM^*(t+s,b)M^*(t+s,c)(f_{B_i,B_j}(b,c) - f_{B_i}(b)f_{B_j}(c))dbdc... \\
   &...\leq  \int_{b=0}^t\int_{c=0}^t\sum_{i=0}^n\sum_{j=0}^n\mathcal{M}^*(t+s)^2f_{B_i,B_j}(b,c)dbdc\\
   &\leq \sum_{i=0}^n\sum_{j=0}^n \int_{b=0}^t\int_{c=0}^t\mathcal{M}^*(t+s)^2(f_{B_i,B_j}(b,c) + f_{B_i}(b)f_{B_j}(c))dbdc\\
   &\leq Z(t,l)^2\mathcal{M}^*(t+s)^2
\end{align}
which gives an overall bound of
\begin{equation}
    \text{var
     }(Z(t+s,l)) \leq  Z(t,l)\mathcal{V}^*(t+s) + Z(t,l)^2\mathcal{M}^*(t+s)^2
\end{equation}
\subsection{Special cases}
To finish, it is helpful to consider a couple of special cases which may arise when the epidemic is large. If the infection times are mostly independent, then
\begin{equation}
    i \neq j \Rightarrow f_{B_i,B_j}(b,c) \sim f_{B_i}(b)f_{B_j}(c)
\end{equation}
while for $i=j$, note that
\begin{align}
   \nonumber&\int_{b=0}^t\int_{c=0}^t\sum_{i=0}^nM^*(t+s,b)M^*(t+s,c)(f_{B_i,B_i}(b,c) - f_{B_i}(b)f_{B_i}(c))dbdc...\\
   &...=\int_{b=0}^t\int_{c=0}^t\sum_{i=0}^nM^*(t+s,b)M^*(t+s,c)(\delta(b-c)f_{B_i}(b) - f_{B_i}(b)f_{B_i}(c))dbdc\\
   &= \int_{b=0}^t \sum_{i=0}^nM^*(t+s,b)^2f_{B_i}(b)db -\sum_{i=0}^n\bigg[\int_bM^*(t+s,b)f_{B_i}(b)db\bigg]^2 
\end{align}
and hence
\begin{equation}
   \text{var}(Z(t+s,l)) \sim \int_{b=0}^t\sum_{i=0}^nV^*(t+s,b)f_{B_i}(b)db+\int_b \sum_{i=0}^nM^*(t+s,b)^2f_{B_i}(b)db -\sum_{i=0}^n\bigg[\int_bM^*(t+s,b)f_{B_i}(b)db\bigg]^2 
\end{equation}
This is still a complicated equation to compute, although the advantage is that one only needs one-dimensional marginal distributions of the infection times, and hence it is significantly more tractable. Moreover, the upper bound on the variance can be improved to
\begin{equation}
    \text{var}(Z(t+s,l)) \leq Z(t,l)\mathcal{V}(t,l) + Z(t,l)\mathcal{M}(t,l)^2
\end{equation}
so that it is proportional to $Z(t,l)$, rather than $Z(t,l)^2$.
\\
\\
\noindent
The simplest case is when the infection times are known - something which may be approximately true if the epidemic is large (and hence has been approximately deterministic in the recent past). In this case, the equation simply reduces to
\begin{equation}
    \text{var}(Z(t+s,l)) \sim\sum_{i=0}^nV^*(t+s,b_i)
\end{equation}
where $b_i$ are the infection times. In this case, the variance can be simply calculated from the quantities $M$ and $V$.
\section{Discrete epidemics}
\subsection{Discrete pgf}
Suppose now that the branching process is entirely discrete (and, for convenience, occurs on integer times). For the lifetime, $L$, of an individual infected at $l$, define
\begin{equation}
    g(u,l) := \mathbb{P}(L=u) \quad \text{and} \quad \overline{G}(u,l) := \mathbb{P}(L \geq u)
\end{equation}
In this discrete setting, it is important to specify exactly inequalities whose strictness is unimportant in the continuous case. In particular, if an individual is infected at time $a$ and has a lifetime of $b$, it will be considered to be infectious at time $a+b$, and will be counted when calculating prevalence at this time. That is, it can infect others at time $a+b$ (and these individuals will be given infection time $a+b$) but will not be able to infect individuals at time $a+b+1$.
\\
\noindent
For the counting process of infections, one can in this case work without a separate infection event process and instead simply use the quantities
\begin{equation}
    q_u(t,l) := \mathbb{P}\bigg(N(t,l)-N(t-1,l) = u\bigg) \quad \text{and} \quad \mathcal{Q}_{(t,l)}(s) := E\bigg(s^{Q(t,l)}\bigg)
\end{equation}
where $Q(t,l)$ has pmf given by $q_u(t,l)$. Hence, each $Q(t,l)$ (which may be zero, unlike $Y$ in the continuous case) gives the number of new infections at time $t$ caused by an individual that was infected at time $l$. Now, note that for $u < t-l$, one has
\begin{equation}
    E\bigg(s^{Z(t,l)}\bigg| L = u\bigg) = E\bigg(s^{\sum_{k=1}^u\sum_{i=1}^{Q(l+k,l)}Z_{ik}(l+u,l)}\bigg)
\end{equation}
where the variables $Z_{ik}$ are iid copies of $Z$. Note that the variables $Q(l+k,l)$ are independent as $N(t,l)$ has indepedent increments, meaning that
\begin{align}
     E\bigg(s^{Z(t,l)}\bigg| L = u\bigg) &= \prod_{k=1}^uE\bigg(s^{\sum_{i=1}^{q(l+k,l)}Z_{ik}(l+u,l+k)}\bigg)\\
     &= \prod_{k=1}^u\mathcal{Q}_{(l+k,l)}\bigg(F(l+u,l+k)\bigg)
\end{align}
Thus, the generating function equation for prevalence can be written as
\begin{equation}
    F(t,l) = s\overline{G}(t-l,l) \prod_{k=1}^{t-l}\mathcal{Q}_{(l+k,l)}\bigg(F(t,l+k)\bigg) + \sum_{u=0}^tg_u\prod_{k=1}^{u}\mathcal{Q}_{(l+k,l)}\bigg(F(t,l+k)\bigg)
\end{equation}
where
\begin{equation}
    \overline{G}(t-l,l) = \mathbb{P}(L \geq t-l)
\end{equation}
The form of the generating function for the discrete case is simpler than the continuous one and might be more amenable to computation.
\subsection{Recovery of the continuous case}
Suppose that each step corresponds to a time interval of $dt << 1$. Suppose further that
\begin{equation}
     \hat{g}(u dt,ldt) dt\sim g_{u,l}, \quad   \hat{t}\sim t dt, \quad\text{and} \quad \hat{l} \sim l dt
\end{equation}
where the quantities with a hat are constant. To ensure continuity in probability, it will be assumed that
\begin{equation}
\hat{q}_u(\hat{t},\hat{l})dt \sim q_u(t,l) \quad \forall u \geq 1 \quad \text{and} \quad q_0(t,l) \sim 1 - \sum_{u=1}^{\infty} dt\hat{q}_u(\hat{t},\hat{l})
\end{equation}
where again, $\hat{q}$ is independent of $dt$. Now, one has
\begin{equation}
    G(t-l,l) = \sum_{u = 0}^{t-l}g_{u,l} \sim \sum_{u = 0}^{\frac{\hat{t} - \hat{l}}{dt}}\hat{g}_{u,l}(udt) dt \sim \int_0^{\hat{t}-\hat{l}}\hat{g}(u,\hat{l}) du := \hat{G}(\hat{t} - \hat{l},l)
\end{equation}
Moreover, one has
\begin{equation}
    \mathcal{Q}_{(t,l)}(s) \sim \bigg(1 - \sum_{u=1}^{\infty}\hat{q}_u(\hat{t},\hat{l})dt\bigg) + \sum_{u=1}^{\infty}s^u\hat{q}_u(\hat{t},\hat{l})dt = 1 + \sum_{u=1}^{\infty}(s^u-1)\hat{q}_u(\hat{t},\hat{l})dt
\end{equation}
Using this relation, setting $\hat{k} := k dt$ and Taylor expanding gives
\begin{align}
    \log\bigg( \prod_{k=1}^{t-l}\mathcal{Q}_{(l+k,l)}(s)\bigg) &\sim \sum_{k=1}^{t-l}\log\bigg(1 + \sum_{u=1}^{\infty}(s^u-1)\hat{q}_u(\hat{l}+\hat{k},\hat{l})dt\bigg)\\
    &\sim \sum_{k=1}^{t-l} \sum_{u=1}^{\infty}(s^u-1)\hat{q}_u(\hat{l}+\hat{k},\hat{l})dt\\
    &\sim \int_{0}^{\hat{t} - \hat{l}}\sum_{u=1}^{\infty}(s^u-1)\hat{q}_u(\hat{l}+\hat{k},\hat{l})d\hat{k}
\end{align}
Hence,
\begin{equation}
\label{eq:discrete_cont_approx}
    F(t,l) \sim (1-\hat{G}(\hat{t}-\hat{l}))\exp\bigg[\int_{0}^{\hat{t} - \hat{l}}\sum_{u=1}^{\infty}(s^u-1)\hat{q}_u(\hat{l}+\hat{k},\hat{l})d\hat{k}\bigg] + \int_0^{t-l}\exp\bigg[\int_{0}^{\hat{u}}\sum_{w=1}^{\infty}(s^w-1)\hat{q}_w(\hat{l}+\hat{k},\hat{l})d\hat{k}\bigg]\hat{g}(\hat{u},\hat{l})d\hat{u}
\end{equation}
It is now possible to define the limiting continuous process. Consider a counting process $N(\hat{t},\hat{l})$ in continuous time with independent increments where infection events occur according to a rate function given by
\begin{equation}
    r(\hat{t},\hat{l}) = \sum_{u=1}^{\infty}\hat{q}_u(\hat{t},\hat{l})
\end{equation}
and where, given that a infection event occurs at $t$ from a particle born at $l$, the infection event is of size $k \geq 0$ with probability 
\begin{equation}
    \frac{\hat{q}_k(\hat{t},\hat{l})}{\sum_{u=1}^{\infty}\hat{q}_u(\hat{t},\hat{l})}.
\end{equation}
Suppose that $J(\hat{t},\hat{l})$ counts the infection events of this process (and hence is an inhomogeneous Poisson Process of rate $r(\hat{t},\hat{l})$) and that $\mathcal{Y}_{(\hat{t},\hat{l})}$ is the generating function of infection event size given that a infection event occurs at $(\hat{t},\hat{l})$. Note that
\begin{equation}
      \int_0^{\hat{t}-\hat{l}}\sum_{u=1}^{\infty}\hat{q}_u(\hat{l}+k,\hat{l})dk =\int_0^{\hat{t}-\hat{l}}r(\hat{l}+k,\hat{l})dk = \mathbb{E}(J(\hat{t},\hat{l}))
\end{equation}
and that
\begin{align}
    \sum_{u=1}^{\infty}s^u\hat{q}_u(\hat{l}+\hat{k},\hat{l}) &=     \sum_{u=1}^{\infty}\bigg(\frac{s^u\hat{y}_u(\hat{l}+\hat{k},\hat{l})}{\sum_{m=1}^{\infty}\hat{y}_m(\hat{l}+\hat{k},\hat{l})}\bigg) \sum_{m=1}^{\infty}\hat{q}_m(\hat{l}+\hat{k},\hat{l})\\
    &= \mathcal{Y}_{(\hat{l}+\hat{k},\hat{l})}(s) \bigg(\sum_{u=1}^{\infty}\hat{q}_u(\hat{l}+\hat{k},\hat{l})\bigg)\\
    &= \mathcal{Y}_{(\hat{l}+\hat{k},\hat{l})}(s)r(\hat{l}+\hat{k},\hat{l})
\end{align}
Hence,
\begin{equation}
    \prod_{k=1}^{t-l}\mathcal{Q}_{(l+k,l)}(s)\sim \exp\bigg[\int_0^{\hat{t}-\hat{l}} r(\hat{l}+k,\hat{l})\mathcal{Y}_{(\hat{l}+k,\hat{l})}(s)dk - \mathbb{E}(J(\hat{t},\hat{l}))\bigg]
\end{equation}
and so, applying this to Supplementary Equation \ref{eq:discrete_cont_approx} shows that the continuous generating function equation is recovered. Note that here, the distribution of $Y$ has been allowed to depend on $l$ (and this is the generating function equation that arises in this case), but the an equation with an $l$-independent $Y$ will arise if the ratio of each $q_k(t,l)$ and $\sum_{k=1}^{\infty}q_k(t,l)$ are independent of $l$.
\subsection{Distinctness from the continuous case}
It is important to note that the relaxation of the assumption that $N$ is continuous in probability necessary in considering the discrete case means that the pgf becomes materially different.

Indeed, one can characterise the discrete case through the continuous framework by imposing that
\begin{equation}
    r(t,l) = \bigg(\sum_{u=1}^{\infty}q_u(t,l)\bigg)\bigg(\sum_{n=1}^{\infty}\delta(l+n-t)\bigg)
\end{equation}
as this is gives probability of $N$ increasing (by whatever number) in the discrete case discussed above. Moreover, again allowing $Y$ to depend on $l$, $Y(t,l)$ has distribution
\begin{equation}
    \mathbb{P}(Y(t,l) = k) = \frac{q_k(t,l)}{\sum_{m=1}^{\infty}q_m(t,l)}
\end{equation}
Now, note that
\begin{equation}
    \lambda(t,l) = \int_l^{t}r(s,l)ds = \sum_{n=1}^{\lfloor t-l\rfloor}\sum_{u=1}^{\infty}q_u(l+n,l)
\end{equation}
where $\lfloor m\rfloor$ denotes the largest integer that is smaller than $m$. Moreover
\begin{equation}
    \int_0^{t-l}\mathcal{Y}_{(l+k,l)}(F(t,l+k))r(l+k,l) = \sum_{n=1}^{\lfloor t-l\rfloor }\sum_{u=1}^{\infty}q_u(l+n,l)\mathcal{Y}_{(l+n,l)}(F(t,l+n))
\end{equation}
We suppose for a contradiction that the pgf in the continuous case is also valid in this discrete setting. Hence (taking $\kappa = 1$)
\begin{align}
    \nonumber F(t,l) &= s(1-G(t-l,l))\exp\bigg[\sum_{n=1}^{\lfloor t-l\rfloor}\sum_{u=1}^{\infty}q_u(l+n,l)\mathcal{Y}_{(l+n,l)}(F(t,l+n)) - \sum_{n=1}^{\lfloor t-l\rfloor}\sum_{u=1}^{\infty}q_u(l+n,l)\bigg]...\\
    &...+\int_0^{t-l}\exp\bigg[\sum_{n=1}^{\lfloor t-l+u\rfloor}\sum_{m=1}^{\infty}q_m(l+n,l)\mathcal{Y}_{(l+n,l)}(F(t,l+n)) - \sum_{n=1}^{\lfloor t-l+u\rfloor}\sum_{m=1}^{\infty}q_m(l+n,l)\bigg]g(u,l)du
\end{align}
Now, note that
\begin{align}
    \mathcal{Y}_{(l+n,l)}(s) &= \sum_{m=1}^{\infty}\frac{s^m q_m(l+n,l)}{\sum_{k=1}^{\infty}q_k(l+n,l)}\\
    &= \frac{1}{\sum_{k=1}^{\infty}q_k(l+n,l)}\bigg(\sum_{m=0}^{\infty}s^m q_m(l+n,l) - q_0(l+n,l)\bigg) \\
    &= \frac{1}{\sum_{k=1}^{\infty}q_k(l+n,l)}\bigg(\mathcal{Q}_{(l+n,l)}(s) - (1 - \sum_{k=1}^{\infty}q_k(l+n,l))\bigg)
\end{align}
and hence
\begin{align}
    &\sum_{n=1}^{\lfloor t-l+u\rfloor}\sum_{m=1}^{\infty}q_m(l+n,l)\mathcal{Y}_{(l+n,l)}(F(t,l+n)) - \sum_{n=1}^{\lfloor t-l+u\rfloor}\sum_{m=1}^{\infty}q_m(l+n,l) \\
    &=\sum_{n=1}^{\lfloor t-l+u\rfloor}\bigg(\mathcal{Q}_{(l+n,l)}(F(t,l+n)) + \sum_{k=1}^{\infty}q_k(l+n,l)\bigg)
\end{align}
which means
\begin{align}
    F(t,l) &= s(1-G(t-l,l))\exp\bigg[\sum_{n=1}^{\lfloor t-l\rfloor}\bigg(\mathcal{Q}_{(l+n,l)}(F(t,l+n)) + \sum_{k=1}^{\infty}q_k(l+n,l)\bigg)\bigg]...\\
    &+\int_0^{t-l}\exp\bigg[\sum_{n=1}^{\lfloor t-l+u\rfloor}\bigg(\mathcal{Q}_{(l+n,l)}(F(t,l+n)) + \sum_{k=1}^{\infty}q_k(l+n,l)\bigg)\bigg]g(u,l)du
\end{align}
Finally, defining $\mathcal{Q}^*(s) := e^{\mathcal{Q}(s)}$ and turning the integral over $g$ into a discrete sum, we have
\begin{equation}
     F(t,l) = s(1-G(t-l,l))\prod_{n=1}^{\lfloor t-l \rfloor}\mathcal{Q}^*(F(t,l+n)) e^{\sum_{k=1}^{\infty}q_k(l+n,l)} + \sum_{u=1}^{\lfloor t-l \rfloor}\prod_{n=1}^{\lfloor t-l+u \rfloor}\mathcal{Q}^*(F(t,l+n)) e^{\sum_{k=1}^{\infty}q_k(l+n,l)} g(u,l)
\end{equation}
This matches very closely with the pgf in the discrete case, but has some extra terms as expected for the contradiction - firstly, the $\mathcal{Q}^*$ in place of the $\mathcal{Q}$, and also the extra $e^{\sum_{k=1}^{\infty}q_k}$ terms. When taking the small $dt$ limit as in the previous subnote, these anomalies disappear, as 
\begin{equation}
    e^{\mathcal{Q}(s)} \sim e^{1 + \alpha dt} \sim 1 + \alpha dt \sim \mathcal{Q}(s)
\end{equation}
and
\begin{equation}
    e^{\sum_{k=1}^{\infty}q_k(l+n,l)} \sim e^{\beta dt} \sim 1
\end{equation}
for some $\alpha$ and $\beta$. Thus, these dissimilarities only appear in the $O(dt^2)$ level (and hence disappear in the small $dt$ limit). However, they will be non-trivial if $dt$ is not small, underlining the importance of the assumption that $N$ is continuous in probability - neglecting such an assumption could lead to materially wrong results in the case of a large step-size.
\subsection{Discrete likelihood}
If the epidemic happens in discrete time, it is significantly easier to calculate the likelihood. Define $A_{k,i}$ to be the number of infections caused at time $k$ by a (still infectious) individual that was infected at time $i$. Then, the number of infections which occur at time $k$ is given by
\begin{equation}
    \mathcal{A}_k(\boldsymbol{y},\boldsymbol{d}) = \sum_{i=0}^{k-1}\sum_{j=1}^{y_i} A^j_{k,i}\mathcal{I}_{\{x_{ij} \leq k\}}  
\end{equation}
where each $A^j_{k,i}$ is an independent copy of $A_{k,i}$ and, similarly to before, $x_{ij}$ is the time at which the jth individual infected at time $i$ stops being infectious. Note that here, as previously in the discrete setting but in contrast to the continuous case, $y_i$ can be zero.
\\
\\
\noindent
Then, the likelihood is simply given by
\begin{equation}
    L(\boldsymbol{y},\boldsymbol{D}) = \bigg(\prod_{k=1}^n\mathbb{P}( \mathcal{A}_k(\boldsymbol{y},\boldsymbol{d}) = y_k) \bigg) \bigg(\prod_{i=1}^n\prod_{j=1}^{y_i}g(x_{ij}-i,i)\bigg)
\end{equation}
where, as we are in the discrete case, $g$ is now a pmf. This gives a log-likelihood of
\begin{equation}
    \ell(\boldsymbol{y},\boldsymbol{D}) = \sum_{k=1}^n\log\bigg(\mathbb{P}( \mathcal{A}_k(\boldsymbol{y},\boldsymbol{d}) = y_k)\bigg) + \sum_{i=1}^n\sum_{j=1}^{y_i} \log(g(x_{ij}-i,i))
\end{equation}

\printbibliography

\end{document}